\documentclass[aps,prfluids,onecolumn,superscriptaddress,longbibliography,nofootinbib,amsmath,amssymb]{revtex4-2}
\usepackage{graphicx}
\usepackage{booktabs}
\usepackage{xcolor}
\usepackage{mathrsfs}
\usepackage{amsthm}
\usepackage[colorlinks=true,linkcolor=blue,citecolor=blue,urlcolor=blue]{hyperref}
\hypersetup{pdftitle={Thermochemical Rossby waves in rapid rotation: from swallowtail geometry to nonlinear selection},pdfauthor={Oleg N. Kirillov}}
\newtheorem{proposition}{Proposition}
\newcommand{\Pran}{\mathit{Pr}}
\newcommand{\Sch}{\mathit{Sc}}
\newcommand{\rd}{\mathrm{d}}
\newcommand{\ii}{\mathrm i}

\begin{document}
\title{Thermochemical Rossby waves in rapid rotation: from swallowtail geometry\texorpdfstring{\\}{ } to nonlinear selection}
\author{Oleg N. Kirillov}
\email{oleg.kirillov@northumbria.ac.uk}
\affiliation{Northumbria University, Newcastle upon Tyne NE1 8ST, United Kingdom}
\begin{abstract}
When heat and solute both destabilise a rapidly rotating fluid, do their buoyancies add? Codensity models assume so, giving a straight onset diagonal; spherical shells show curved and kinked thresholds. We study the Busse--Simitev rotating annulus, whose normal modes obey an exact complex cubic dispersion relation. Combining the Bilharz criterion with the neutral-curve envelope yields, in the rapid-rotation limit, an explicit swallowtail wave front, in Arnold's sense, that bends the codensity diagonal into smooth and then kinked thresholds as the diffusivity ratio decreases. Its cusp edges lie inside the unstable domain and its exposed kink ends at the swallowtail singularity; these are singularities of wave-number selection, without coalescence of temporal eigenvalues, and the swallowtail diffusivity ratio has analytical small- and large-Prandtl-number limits. At a kink, broad, fast thermal and narrow, slow compositional Rossby waves become neutral together; hidden arcs mark the birth and merger of unstable wave-number bands. The same geometry enters nonlinear selection: mean-scalar feedback factors through neutral-arc normals and scalar responses, so the mean-profile coexistence margin is positive at every corner of the leading front and vanishes quadratically with the kink angle as the waves merge. Complete cubic coefficients predict robust coexistence at a reference corner and a narrow bistable interval near merger. Periodic-cell simulations away from merger confirm the coexisting waves and their invasion thresholds.
\end{abstract}
\maketitle
\section{Introduction}\label{sec:introduction}
\begin{flushright}8a4cd9ef7i3l4m6n4opqq7rss5t14u\footnote{\emph{Data curvarum familia quotcunque parametros involvente, involucrum invenire; et vice versa}: given a family of curves depending on any number of parameters, to find the envelope, and conversely. After the anagram in which Newton concealed the calculus in his letter of 24 October 1676 to Oldenburg for Leibniz.}\end{flushright}
Convection in planetary cores is driven both by heat and by the light elements released as the inner core grows. Both make the fluid lighter, but heat diffuses orders of magnitude faster than composition, so their effects need not add. In rapid rotation this convection sets in as thermal Rossby waves, columns aligned with the rotation axis that drift around it.

Many models nevertheless combine the two sources into a single codensity with an effective diffusivity \citep{Braginsky1995}. When the two scalars have coincident normalised background profiles and boundary conditions, the sources become interchangeable, and the critical Rayleigh numbers lie on a straight diagonal in the plane of the two Rayleigh numbers. Calculations that keep the two diffusivities separate tell a different story, with non-additive thresholds and abrupt changes in the scale and drift of convection in rotating spherical shells \citep{Trumper2012,Tassin2021}. When the Prandtl and Schmidt numbers are equal, the threshold is indeed a straight line \citep{Silva2019}. As the diffusivities move apart, the line bends and then develops a kink between a thermal and a compositional family of modes \citep{Net2012,Silva2019,Mather2021}. Across this kink the critical azimuthal wave number found by \citet{Net2012} jumps from 3 to 6 while the drift frequency drops 25-fold, and in the shell of \citet{Mather2021} the wave number jumps from 9 to 25.

Such thresholds are found by computing a neutral curve for each azimuthal wave number and taking the lowest, so the kink appears where two families of curves cross. This locates the kink but leaves open why it forms, at which diffusivity ratio it first appears, which geometry turns a straight threshold into a kinked one, and what the two waves do once both grow. In single-scalar convection near mode crossings, both coexistence and bistability occur, with double columns and bistability in a rotating annulus \citep{Pino2001} and bistable thermal waves with unstable mixed tori in a spherical shell \citep{Sanchez2013}. \citet{Busse1986} found mixed states of two radial modes of one scalar, and weakly nonlinear analyses of spherical shells show that zonal flows can make the onset subcritical \citep{Skene2025}. To our knowledge, however, the interaction of the two waves at the kink of a two-scalar threshold has not been analysed.

A model that keeps both diffusivities and still admits analysis is the small-gap rotating annulus with conical end caps \citep{Busse1986,Busse2002}. \citet{Simitev2011} extended it to double-diffusive convection and studied the onset and finite-amplitude convection under pure and mixed driving. Its normal modes obey an exact cubic dispersion relation, which also appears in the analysis of stably stratified cores by \citet{Monville2019}. Because the relation is cubic, the onset problem becomes algebraic, and its linear responses serve as the building blocks of the two-wave interaction.

We therefore find the onset threshold as an envelope. At a fixed wave number, the Bilharz criterion \citep{Bilharz1944} gives the stability conditions of the complex cubic, and their boundary is a neutral curve. The onset threshold is the part of the envelope of these curves that bounds the region where every wave number is stable. The same construction located onset and the critical modes in visco-diffusive swirling flows \citep{KirillovMutabazi2024,KirillovMutabazi2025}, where the envelope surface formed a pleat with two folds, the geometry of a cusp catastrophe (Fig.~3 of Ref.~\cite{KirillovMutabazi2024}). Reference~\cite{Kirillov2025} reviews envelope thresholds of rotating visco-diffusive flows. With the diffusivity ratio as a parameter, the envelope becomes a wave front in the sense of \citet{Arnold1990}, whose singularities follow the classification of wave fronts. Here the next singularity of that hierarchy, the swallowtail, organises the kink.

In the rapid-rotation limit the envelope is explicit. As the diffusivity ratio decreases, the codensity diagonal bends smoothly and then, below a swallowtail value, develops a kink where a broad, fast thermal wave and a narrow, slow compositional wave become neutral together. The swallowtail value depends algebraically on the Prandtl number and has closed-form limits, so a kink in the annulus requires a Lewis number above a critical value between $5.26$ and $6.16$. The cusps and the swallowtail are singularities of wave-number selection rather than coalescences of eigenvalues, and the hidden arcs of the front mark where bands of unstable wave numbers are born and merge. The same geometry enters the nonlinear interaction, because the coupling of the two waves through the mean scalar profiles factorises into the normals of the two neutral arcs and the responses of the two scalars. The resulting coexistence margin is positive at every leading-order corner and closes quadratically with the kink angle as the waves merge. Complete cubic coefficients confirm coexistence at a reference corner and predict a narrow bistable interval near the merger. Away from the merger, simulations in a periodic cell confirm the coexisting waves and their invasion thresholds.

Section~\ref{sec:model} introduces the annulus and its cubic, and Sec.~\ref{sec:envelope} explains why the onset threshold is an envelope. Section~\ref{sec:swallowtail} derives the swallowtail and the diffusivity ratio at which the kink appears, and Sec.~\ref{sec:front} reads the front as a map of the unstable wave numbers. Section~\ref{sec:nonlinear} derives the coexistence criterion from the same geometry, and Sec.~\ref{sec:finite} follows the interaction near the swallowtail at finite rotation. Sections~\ref{sec:validation} and~\ref{sec:sphere} test the predictions in simulations and compare them with spherical calculations, and Sec.~\ref{sec:conclusions} concludes. The appendices collect the conventions, proofs, algebra and numerical methods.

\section{Two buoyancies in a rotating annulus}\label{sec:model}
Consider fluid between two coaxial cylinders of mean radius $s_0$ and gap $d\ll s_0$ that rotate rigidly with angular velocity $\Omega$ about their axis $z$. The walls hold fixed temperatures and concentrations, and the centrifugal acceleration acts as gravity. Conical caps of small slope $\eta_0$ close the gap axially \citep{Busse1986,Busse2002}. Rapid rotation makes the motion nearly uniform along $z$, so the flow consists of columns parallel to the axis. A column that moves radially is stretched or compressed beneath the sloping caps, and conservation of vorticity turns this into the topographic $\beta$-effect, with $\beta=4\eta_0\Omega d^3/(\nu h)>0$ for axial length $h$. Disturbances therefore propagate around the annulus as Rossby waves. We scale time by $d^2/\nu$ and use local coordinates $x=(s-s_0)/d+\frac12$ and $y=s_0\varphi/d$, increasing prograde, for radius $s$ and azimuth $\varphi$.

With kinematic viscosity $\nu$ and thermal and compositional diffusivities $\kappa_T$, $\kappa_C$, the Prandtl number is $\Pran=\nu/\kappa_T$, the diffusivity ratio is $\tau=\kappa_C/\kappa_T$ and the Schmidt number is $\Sch=\Pran/\tau$. The thermal and compositional Rayleigh numbers $R$ and $S$ use their respective scalar diffusivities and are positive for destabilising gradients (Appendix~\ref{sec:conventions}). Deviations $\psi$, $\Theta$, $\Xi$ of stream function, temperature and concentration from rigid rotation with linear conductive profiles obey \citep{Simitev2011}
\begin{align}
 (\partial_t-\nabla^2)\nabla^2\psi+J(\psi,\nabla^2\psi)-\beta\partial_y\psi
       +\partial_y(R\Theta+\tau S\Xi)&=0,\nonumber\\
 \Pran[\partial_t\Theta+J(\psi,\Theta)]+\partial_y\psi-\nabla^2\Theta&=0,\label{eq:nl}\\
 \Pran[\partial_t\Xi+J(\psi,\Xi)]+\partial_y\psi-\tau\nabla^2\Xi&=0,\nonumber
\end{align}
where the Jacobian $J(a,b)=a_xb_y-a_yb_x$ gives advection by the velocity $(-\psi_y,\psi_x)$. Stress-free walls with fixed scalar values give $\psi=\partial_x^2\psi=\Theta=\Xi=0$ at $x=0,1$. The first equation is the vorticity balance: viscous diffusion, the $\beta$-effect and the torques of the two buoyancies. The other two equations carry temperature and concentration, and they differ only in the diffusivity, $1$ against $\tau$.

For infinitesimal deviations the Jacobians drop out. Normal modes
\begin{equation*}
(\psi,\Theta,\Xi)^{\mathsf T}=(\hat\psi,\hat\Theta,\hat\Xi)^{\mathsf T}\sin(kx)\,e^{\lambda t+\mathrm{i}ly},
\end{equation*}
with $k=n\pi$, radial index $n\geq1$, azimuthal wave number $l$ and $H=k^2+l^2$, satisfy $M(\lambda)(\hat\psi,\hat\Theta,\hat\Xi)^{\mathsf T}=0$ with
\begin{equation}
 M(\lambda)=\begin{pmatrix}
 H(\lambda+H)+\mathrm{i}\beta l&-\mathrm{i}lR&-\mathrm{i}\tau lS\\
 \mathrm{i}l&\Pran\,\lambda+H&0\\ \mathrm{i}l&0&\Pran\,\lambda+\tau H
 \end{pmatrix}.
 \label{eq:M}
\end{equation}
The eigenvalue $\lambda$ has growth rate $\mathrm{Re}\,\lambda$ and frequency $\omega=\mathrm{Im}\,\lambda$. Periodicity around the ring quantises the wave number, $l=m\Delta l$ with integer $m$ and $\Delta l=d/s_0$; since $d\ll s_0$ we treat $l$ as continuous.

Eliminating the two scalar amplitudes gives the dispersion relation. Each scalar responds to the stream function through its own transfer function,
\begin{equation*}
T_T=(\Pran\,\lambda+H)^{-1},\qquad T_C=(\Pran\,\lambda+\tau H)^{-1},
\end{equation*}
and the vorticity balance becomes $\mathcal F(\lambda)=0$ with
\begin{equation*}
\mathcal F(\lambda)=H(\lambda+H)+\mathrm{i}\beta l-l^2(RT_T+\tau ST_C).
\end{equation*}
Multiplying by the two denominators gives the cubic of \citet{Simitev2011},
\begin{equation}
 f(\lambda)=(\Pran\,\lambda+H)(\Pran\,\lambda+\tau H)\,\mathcal F(\lambda)=\sum_{j=0}^3 f_j\lambda^j,\label{eq:cubic}
\end{equation}
which is $\tau$ times the cubic (C12) of \citet{Monville2019}. The two prefactors vanish only at $\mathrm{Re}\,\lambda<0$, so $f$ and $\mathcal F$ share their growing and neutral roots. The transfer functions carry the physics of the problem. The same flow drives both scalars, but the solute, which diffuses more slowly, responds more strongly and with a larger phase lag.

The exchange $(\Pran,\tau,R,S)\mapsto(\Pran/\tau,1/\tau,S,R)$ only rescales $f$ by $\tau^{-2}$: renaming the two scalars changes nothing. Unequal diffusivities can therefore be taken with $\tau<1$, heat diffusing faster.

The first onset in the continuous-$l$ problem has radial index $n=1$. Take a neutral mode with $k>\pi$ and choose $\zeta>1$ with $\zeta H-\zeta^4l^2=\pi^2$; such a $\zeta$ exists because the left side equals $k^2>\pi^2$ at $\zeta=1$ and vanishes at $\zeta=(H/l^2)^{1/3}$. The transformation
\begin{equation*}
(\lambda,l,H,R,S)\mapsto(\zeta\lambda,\zeta^2l,\zeta H,R/\zeta,S/\zeta)
\end{equation*}
multiplies \eqref{eq:cubic} by $\zeta^4$ at fixed $\beta$, $\Pran$ and $\tau$: each prefactor gains $\zeta$, and $\mathcal F$ gains $\zeta^2$. The neutral root $\mathrm{i}\omega$ thus becomes the neutral root $\mathrm{i}\zeta\omega$ of an $n=1$ mode in the same annulus, at the same ratio $S/R$ but with weaker forcing. Every neutral mode with $n\geq2$ is preceded by a neutral mode with $n=1$, and we set $k=\pi$ from now on.

When the diffusivities are equal, the two scalars act as one. At $\tau=1$ the cubic factorises,
\begin{equation*}
f=(\Pran\,\lambda+H)\{[H(\lambda+H)+\mathrm{i}\beta l](\Pran\,\lambda+H)-l^2(R+S)\},
\end{equation*}
and onset depends on $R+S$ alone: the threshold is the straight codensity diagonal. The topographic term decides what happens when $\tau\neq1$. On the imaginary axis,
\begin{equation*}
 \mathrm{Im}\,\mathcal F(\mathrm{i}\omega)=\beta l+\omega\Bigl[H+\Pran\,l^2\Bigl(\frac{R}{H^2+\Pran^2\omega^2}+\frac{\tau S}{\tau^2H^2+\Pran^2\omega^2}\Bigr)\Bigr],
\end{equation*}
and the bracket is positive for $R,S\ge0$. With the topographic term set to zero, $\beta=0$, every neutral mode is therefore stationary, and $\mathcal F(0)=H^2-l^2(R+S)/H$ makes the neutral condition $R+S=H^3/l^2$ for every $\tau$: the threshold is again a straight line, as in the classical thermohaline problem with the same boundary conditions for both scalars \citep{Nield1967,Tuckerman2001}. The topographic drift makes the two scalar responses differ in magnitude and phase when $\tau\neq1$, and this bends the threshold; Sec.~\ref{sec:swallowtail} shows when it kinks. The rest of this paper asks what happens to this line when $\tau<1$, and what the waves do beyond it.

\section{Many neutral curves, one threshold}\label{sec:envelope}
For each wave number, the cubic \eqref{eq:cubic} decides whether the corresponding wave grows. The onset threshold is a different object: it bounds the set of forcings at which no wave grows. This section constructs it and shows where its singularities come from.

We take $l>0$: the mode $l=0$ is diffusive and decays, and negative $l$ follow by complex conjugation. Write $f_j=\alpha_j+\mathrm{i}b_j$, with $\alpha_3>0$ and $b_3=0$. By the Bilharz criterion \citep{Bilharz1944}, all roots of the complex cubic have negative real parts if and only if the leading principal minors $D_1$, $D_2$ and $D_3$ of orders 2, 4 and 6 of
\begin{equation}
 \mathsf B=\begin{pmatrix}
 \alpha_0&-b_0&0&0&0&0\\
 b_1&\alpha_1&\alpha_0&-b_0&0&0\\
 -\alpha_2&b_2&b_1&\alpha_1&\alpha_0&-b_0\\
 0&-\alpha_3&-\alpha_2&b_2&b_1&\alpha_1\\
 0&0&0&-\alpha_3&-\alpha_2&b_2\\
 0&0&0&0&0&-\alpha_3
 \end{pmatrix}\label{eq:B}
\end{equation}
are positive. When all minors are non-zero, the sign changes in $(1,D_1,D_2,D_3)$ count the growing roots \citep{Bilharz1944,Kirillov2021}. With
\begin{equation*}
\mathcal E=b_2\alpha_0-\alpha_2b_0,\qquad
\mathcal K=\alpha_2\alpha_1+b_2b_1-\alpha_3\alpha_0,
\end{equation*}
the minors satisfy
\begin{equation}
 D_2=\mathcal K D_1-\mathcal E^2,
 \qquad
 D_3\big|_{D_2=0}=-\frac{\alpha_3^2}{D_1}
           (b_1D_1-\alpha_0\mathcal E)^2\quad(D_1>0).
 \label{eq:minor}
\end{equation}
These identities show which minor fails first. Since $\alpha_2,b_2,b_0>0$, approaching $D_1=0$ from stability would require $\mathcal E=0$, hence $\alpha_0>0$ and $\alpha_2\alpha_1+b_2b_1=(b_2/b_0)D_1=0$, so that $\mathcal K<0$, which contradicts $D_2>0$. On $D_2=0$ with $D_1>0$, equation \eqref{eq:minor} gives $D_3\leq0$, so $D_3$ vanishes first. Every stability boundary at fixed $l$ therefore lies on $D_3=0$, where one wave becomes neutral (Appendix~\ref{sec:selection-boundary}).

The highest minor $D_3$ is proportional to the resultant of the real and imaginary parts of $f(\mathrm{i}\omega)$, so, for $R,S>0$, $D_3=0$ is equivalent to the neutral conditions (Appendix~\ref{app:resultant})
\begin{equation*}
\mathrm{Re}\,f(\mathrm{i}\omega)=\mathrm{Im}\,f(\mathrm{i}\omega)=0.
\end{equation*}
They are affine in $R$ and $S$. With $H=\pi^2+l^2$, the frequency parameter $X=\Pran\,\omega/H$, $a=1+1/\Pran$ and $b=1+\tau/\Pran$, they give the neutral curve of the wave number $l$ in parametric form,
\begin{equation}
 R(X,l)=\frac{H(X^2+1)(bH^2X+\tau\beta l)}{(1-\tau)l^2X},\qquad
 S(X,l)=-\frac{H(X^2+\tau^2)(aH^2X+\beta l)}{(1-\tau)\tau l^2X}.
 \label{eq:family}
\end{equation}
In the quadrant $R,S>0$ the frequency parameter is confined to
\begin{equation*}
-\frac{\beta l}{aH^2}<X<-\frac{\tau\beta l}{bH^2},
\end{equation*}
so for $l>0$ every neutral wave drifts prograde, $\omega<0$. The same parameter fixes the phases of the scalar responses: at $\lambda=\mathrm{i}\omega$,
\begin{equation*}
T_T=\frac{1}{H(1+\mathrm{i}X)},\qquad T_C=\frac{1}{H(\tau+\mathrm{i}X)}.
\end{equation*}

The neutral curve \eqref{eq:family} is not yet the onset threshold. Where it crosses the region in which waves of other wave numbers already grow, the motionless state is unstable although the wave $l$ is neutral. The state is stable only if every admissible wave number is stable, so its stable set is the intersection of the fixed-$l$ stable sets, and the onset threshold is the boundary of that intersection. Along a ray in the $(R,S)$ plane, onset is the smallest forcing at which some wave becomes neutral, and at that minimum the neutral forcing is stationary in $l$. The onset threshold is therefore the exposed part of the envelope of the family \eqref{eq:family} over $l$. The envelope is where the parametrisation folds,
\begin{equation}
 \frac{\partial(R,S)}{\partial(X,l)}=0,
 \label{eq:envelope}
\end{equation}
which is a quartic equation in $X$ (Appendix~\ref{sec:quartic}). Differentiating $D_3(R(X,l),S(X,l);l)=0$ on a regular sheet identifies \eqref{eq:envelope} with
\begin{equation*}
D_3=(\partial_lD_3)_{R,S}=0,
\end{equation*}
the Bilharz envelope: neutral growth is stationary in wave number. Each envelope point retains its preimage $(X,l)$, hence the wave number and the frequency $\omega=HX/\Pran$ of the wave that becomes neutral there. The parts of the envelope that are not exposed lie inside the unstable region; Sec.~\ref{sec:front} shows that they mark changes in the set of unstable wave numbers.

The singularities of this envelope have a simple origin. At a neutral wave with $l>0$ in $R,S>0$, direct differentiation of $\mathcal F$ gives
\begin{equation}
 \operatorname{Im}\mathcal F_\lambda
 =-\frac{2\Pran\,l^2X}{H^2}
   \left[\frac{R}{(1+X^2)^2}
    +\frac{\tau^2S}{(\tau^2+X^2)^2}\right]>0,
 \label{eq:neutral-root-simplicity}
\end{equation}
because $X<0$. Neither prefactor of \eqref{eq:cubic} vanishes on the imaginary axis, so $f_\lambda\neq0$: every neutral root is simple. \citet{Tuckerman2001} found an avoided crossing between real growth rates at a fixed wave number in nonrotating convection with cooperating gradients. Here, by contrast, the threshold kinks where two simple neutral waves with different wave numbers reach onset together. The temporal eigenvalue never degenerates on the neutral surface, and every singularity of the threshold comes from the selection of the wave number. With $\sigma(l;R,S)=\max_j\mathrm{Re}\,\lambda_j$ the largest growth rate at wave number $l$, on the set $\sigma=\sigma_l=0$ the normal to the neutral surface lifts the envelope to a Legendrian surface wherever $(\sigma,\sigma_l)$ has rank two in $(l,R,S,\tau)$. With $\tau$ as a parameter, the envelope is therefore a wave front in the sense of \citet{Arnold1990}, and its singularities are those of wave fronts.

The two sources of singularity differ. In visco-diffusive azimuthal magnetorotational instability, the neutral surface self-intersects at an angle that closes at Whitney umbrella points, where eigenvalues at fixed wave number collide \citep{Kirillov2017}, a singularity of matrix families depending on parameters \citep{Arnold1971}; see also Sec.~2 and Fig.~3 of Ref.~\cite{Kirillov2025}. Here every neutral root is simple, and the kink and its swallowtail come from wave-number selection alone.

\section{The swallowtail}\label{sec:swallowtail}
\begin{figure}[!htbp]
 \centering\includegraphics[width=\textwidth]{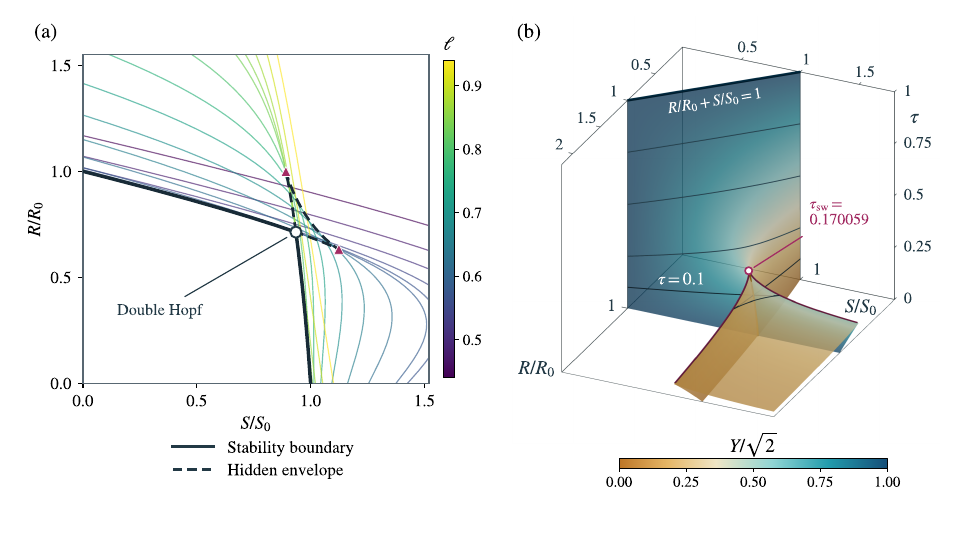}
 \caption{Neutral curves and their swallowtail surface for $\Pran=0.3$. (a) Leading neutral curves \eqref{eq:lead} at $\tau=0.1$, coloured by $\ell=l/\beta^{1/3}$, and their envelope \eqref{eq:radical}; triangles mark the cusps. (b) The full envelope surface in $(R/R_0,S/S_0,\tau)$, with $\tau=0$ a one-sided limit. The codensity diagonal at $\tau=1$ turns into smoothly bent and then kinked sections; magenta cusp edges and the amber corner locus meet at the labelled swallowtail point, and the labelled $\tau=0.1$ section is the geometry of (a). Surface colours give $Y/\sqrt2$, where $Y=-\Pran\,\omega/H>0$. Both panels use the rapid-rotation limit and the thresholds $R_0$, $S_0$ of \eqref{eq:pure}. Plotted with the AI-assisted scripts described in Appendix~\ref{sec:numerics}.}
 \label{fig:front}
\end{figure}

In the rapid-rotation limit the envelope can be written explicitly. This section shows that, as $\tau$ decreases, it turns the codensity diagonal into a kinked threshold through a swallowtail (Fig.~\ref{fig:front}), and it finds the diffusivity ratio at which the kink appears.

Let $\beta\to\infty$ at fixed $\Pran$, $\tau$ and $k=\pi$, with $l=\beta^{1/3}\ell$ and $(R,S)=\beta^{4/3}(\hat R,\hat S)$. Heat alone and solute alone become unstable at
\begin{equation}
 R_0=3\left[\frac{\Pran\,\beta}{\sqrt2(1+\Pran)}\right]^{4/3},\qquad
 S_0=3\left[\frac{\Sch\,\beta}{\sqrt2(1+\Sch)}\right]^{4/3},
 \label{eq:pure}
\end{equation}
with wave numbers $\ell_{T,0}^3=\Pran/[\sqrt2(1+\Pran)]$ and $\ell_{C,0}^3=\Sch/[\sqrt2(1+\Sch)]$. These are the single-scalar balances of rotation and diffusion. The larger Schmidt number gives narrower columns, and the slower compositional response gives slower drift. The line $R/R_0+S/S_0=1$ is then only an additive reference: it is where the threshold would lie if the two buoyancies simply added.

Let $\epsilon=k^2\beta^{-2/3}$ and $Y=-X>0$. Then $H=\beta^{2/3}(\ell^2+\epsilon)$, and finite-rotation corrections enter through $\epsilon/\ell^2=k^2/l^2=O(\beta^{-2/3})$: the columns are thin, $l\gg k$. At $\epsilon=0$ the family \eqref{eq:family} becomes
\begin{equation}
 \hat R(Y,\ell)=\frac{\ell(Y^2+1)(b\ell^3Y-\tau)}{(1-\tau)Y},\qquad
 \hat S(Y,\ell)=\frac{\ell(Y^2+\tau^2)(1-a\ell^3Y)}{(1-\tau)\tau Y}.
 \label{eq:lead}
\end{equation}

\begin{proposition}[global radical]\label{prop:radical}
Set $w=\ell^3Y$ and $Z=Y^2$. The Jacobian of \eqref{eq:lead} vanishes exactly when $Z^2-\mathcal C(w)Z-2\tau^2=0$. Its unique positive root is the global radical
\begin{align}
 \mathcal C(w)&=4(1+\tau)abw+\tau(1+\tau)/w-(2+\tau)(1+2\tau)-5\tau(1+\tau)/\Pran,\nonumber\\
 Z(w)&=\tfrac12[\mathcal C(w)+(\mathcal C(w)^2+8\tau^2)^{1/2}],\label{eq:radicalZ}
\end{align}
and it parametrises the complete leading envelope in $R>0$, $S>0$:
\begin{align}
 \hat R(w)&=\frac{w^{1/3}(Z+1)(bw-\tau)}{(1-\tau)Z^{2/3}},\qquad
 \hat S(w)=\frac{w^{1/3}(Z+\tau^2)(1-aw)}{(1-\tau)\tau Z^{2/3}},\label{eq:radical}\\
 \ell&=(w/\sqrt Z)^{1/3},\qquad X=-\sqrt Z,\qquad \tau/b\leq w\leq1/a.\nonumber
\end{align}
\end{proposition}

\emph{Proof.} The map $(\ell,Y)\mapsto(w,Z)$ is a diffeomorphism for $\ell,Y>0$, and in the new variables
\begin{equation}
 \frac{\partial(\hat R,\hat S)}{\partial(w,Z)}
 =\frac{w^{2/3}}{3\tau(1-\tau)Z^{7/3}}
       [Z^2-\mathcal C(w)Z-2\tau^2],
 \label{eq:radical-certificate}
\end{equation}
so the envelope is the zero set of the bracket. The product of its two roots is $-2\tau^2<0$, so exactly one root is positive, and a single square root covers the whole curve. Positivity of $\hat R$ and $\hat S$ fixes $\tau/b\leq w\leq1/a$, and at the two ends $Z=2\tau^2$ and $Z=2$ give the pure compositional and thermal onsets \eqref{eq:pure}. $\square$

The ends of the envelope already show why two waves compete. Heat alone is most effective at the frequency parameter $X=-\sqrt2$, and solute alone at $X=-\sqrt2\,\tau$; for $\tau\ll1$ these are very different waves.

Cusps of the envelope satisfy $\rd\hat R/\rd w=\rd\hat S/\rd w=0$. A corner is a pair $w_T\neq w_C$ with equal $(\hat R,\hat S)$ at which all other admissible modes are stable; there two waves become neutral together. At $\Pran=0.3$ and $\tau=0.1$ the envelope has two cusps, at $w=0.1214$ and $0.1622$, and a self-intersection (Fig.~\ref{fig:front}(a)). Its exposed arcs meet at $R_\times/R_0=0.7130$, $S_\times/S_0=0.9355$, with thermal and compositional preimages $(\ell_T,X_T)=(0.5701,-1.014)$ and $(\ell_C,X_C)=(0.8468,-0.1634)$. Both waves are prograde. The thermal wave is broader and faster: its frequency is $2.81$ times and its phase speed $4.18$ times that of the compositional wave. The sum $R_\times/R_0+S_\times/S_0=1.6485$, rather than $1$, measures the non-additivity.

Across the kink, the wave that grows first changes from a broad, fast thermal wave to a narrow, slow compositional wave, and at the kink both are neutral. A mixed forcing can drive either wave, but each scalar works best at its own frequency, so onset switches from one arc of the envelope to the other and the selected width and drift jump at the corner. In spherical shells at Ekman number $E=10^{-3}$, $\Pran=0.1$ and $\tau=0.01$, the switch raises the azimuthal wave number from 3 to 6 and cuts the drift frequency 25-fold \citep{Net2012}; at the same $\Pran$ and $\tau$ the leading corner gives $\ell_C/\ell_T=2.25$ and $|\omega_T/\omega_C|=19.6$ (Sec.~\ref{sec:sphere}). At $\beta=3.786\times10^6$, the value given by the annulus calibration of \citet{Monville2019} at Ekman number $10^{-6}$, the exact cubic \eqref{eq:cubic} with Bilharz selection reproduces the leading corner to within $0.06\%$ (Appendix~\ref{sec:leading-surface}); we call this finite-rotation corner the reference corner.

Varying $\tau$ sweeps the fronts into the surface of Fig.~\ref{fig:front}(b). Its cusp edges and its corner locus meet at the swallowtail point $(R/R_0,S/S_0,\tau_{\rm sw})=(0.5738,0.8323,0.17006)$. As $\tau$ decreases from $1$, the straight threshold first bends smoothly. Below $\tau_{\rm sw}$ it develops hidden folds and a kinked exposed boundary, because the diffusivities now differ enough for heat and solute to select distinct waves. Each fixed-$\tau$ corner is a double-Hopf point with simple roots at distinct wave numbers; at the swallowtail the two waves merge.

\subsection{Anatomy of the kink}\label{sec:normal-form}
At the swallowtail point, the scaled growth rate $F=\hat\sigma=\sigma/\beta^{2/3}$, as a function of $\ell$ at given $(\hat R,\hat S,\tau)$, has a degenerate maximum in the wave number,
\begin{equation*}
F=F_\ell=F_{\ell\ell}=F_{\ell\ell\ell}=0>F_{\ell\ell\ell\ell}.
\end{equation*}
Two conditions make this degeneracy generic. The three parameters $(\hat R,\hat S,\tau)$ must unfold it, so the Jacobian $J_{\rm unf}$ of $(F,F_\ell,F_{\ell\ell})$ with respect to them is non-singular. The physical slices $\tau=\mathrm{const}$ must cut it transversally, so the minor $B_\tau$ of $(F,F_\ell)$ with respect to $(\hat R,\hat S)$ is non-zero; this is the time-function condition of \citet[\S~3.3]{Arnold1990}, with the diffusivity ratio playing the role of time. Both conditions hold at $\Pran=0.3$ (Appendix~\ref{sec:leading-surface}); they are evaluated there in the logarithmic variables $(\ln\hat R,\ln\hat S,\ln\tau)$, which differ from $(\hat R,\hat S,\tau)$ by a diagonal rescaling and so do not affect whether the determinants vanish. The front is then locally the front of the generating family
\begin{equation*}
\mathcal Q(u;\delta_*,\xi,\eta)=u^4-\delta_*u^2+\xi u+\eta,\qquad \mathcal Q=\mathcal Q_u=0
\end{equation*}
\citep[\S~3.2]{Arnold1990}. Here $u$ is a local wave-number coordinate, $(\xi,\eta)$ are local coordinates in the Rayleigh plane, and $\delta_*$ is proportional to $\tau_{\rm sw}-\tau$. In a section $\delta_*>0$,
\begin{align}
 \xi&=-4u^3+2\delta_*u,\qquad \eta=3u^4-\delta_*u^2,\nonumber\\
 \text{cusps:}\quad u_\pm&=\pm\sqrt{\delta_*/6},\quad
 \xi_\pm=\pm\frac{4\delta_*^{3/2}}{3\sqrt6},\quad \eta_\pm=-\delta_*^2/12,\nonumber\\
 \text{corner:}\quad (\xi,\eta)&=(0,\delta_*^2/4),\qquad u=\pm\sqrt{\delta_*/2}.
 \label{eq:sw-section-normal-form}
\end{align}
The normal form gives the laws of the kink. The cusp separation scales as $\delta_*^{3/2}$, the distance of the corner from the chord of the cusps as $\delta_*^2$, and the separation of the two wave numbers and the kink angle as $\delta_*^{1/2}$. In these coordinates the distance is $\delta_*^2/3$ and the angle is $2\arctan\sqrt{\delta_*/2}$, and a smooth change of coordinates preserves the exponents. The swallowtail is therefore the generic way in which a straight threshold acquires a kink as one parameter varies. An ordinary cusp, by contrast, is never exposed: near its wave number $l_*$ the growth rate behaves as $\sigma\approx\sigma_{lll}(l-l_*)^3/6$ and changes sign, so waves on one side already grow.

\begin{figure}[!htbp]
 \centering\includegraphics[width=.66\textwidth]{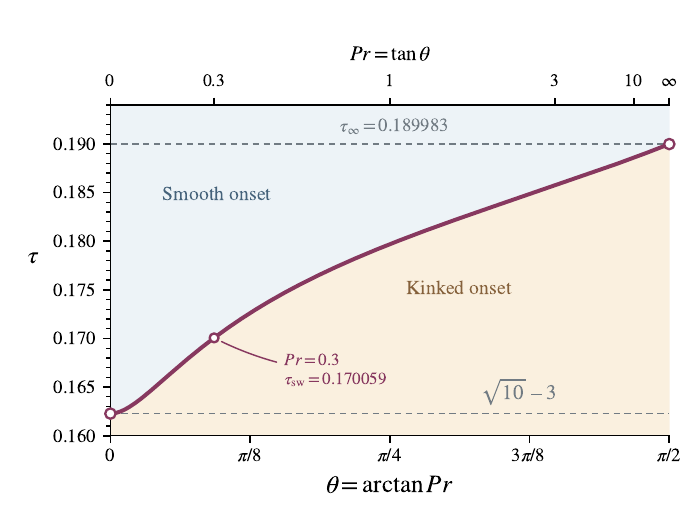}
 \caption{Swallowtail locus $\tau_{\rm sw}(\Pran)$ of the leading rapid-rotation front of the annulus, compactified by $\Pran=\tan\theta$ (upper axis: $\Pran$). Dashed lines are the analytical limits $\sqrt{10}-3$ and $\tau_\infty$ of \eqref{eq:sw-endpoint-limits}, and the open endpoints are one-sided limits. Below the numerically continued curve the onset is kinked and two waves compete; above it the onset is smooth. The marked point is $\Pran=0.3$, $\tau_{\rm sw}=0.1700590$. In the annulus, a kink therefore requires a Lewis number $Le=1/\tau$ above $Le_{\rm sw}=1/\tau_{\rm sw}$, which lies between $5.26$ and $6.16$. Plotted with the AI-assisted scripts described in Appendix~\ref{sec:numerics}.}
 \label{fig:sw-locus}
\end{figure}

\subsection{When the kink appears}\label{sec:locus}
The swallowtail diffusivity ratio $\tau_{\rm sw}(\Pran)$ is algebraic. The cusp condition and the merger of two cusps are polynomial equations in the front variables; eliminating them leaves one polynomial equation $\Phi=0$ in $\tau$ and $\Pran$ (Appendix~\ref{sec:sw-locus}). The elimination equation defines a unique, increasing algebraic branch $\tau_{\rm sw}(\Pran)$ (Appendix~\ref{sec:sw-locus}), with the closed-form limits
\begin{equation}
\begin{aligned}
 \tau_{\rm sw}&=\sqrt{10}-3+\frac{9(10-\sqrt{10})}{200}\Pran^2+O(\Pran^3),
 &&\Pran\to0,\\
 \tau_{\rm sw}&=\tau_\infty-\frac{C_\infty}{\Pran}+O(\Pran^{-2}),
 &&\Pran\to\infty,
\end{aligned}
\label{eq:sw-endpoint-limits}
\end{equation}
where $\tau_\infty=0.18998$ is fixed by the root of a sextic and $C_\infty=0.01453$. Figure~\ref{fig:sw-locus} shows the branch over the full range of $\Pran$; numerical continuation traces the transition from smooth to kinked onset along it. In the annulus, a kink therefore requires a Lewis number $Le=1/\tau$ above $Le_{\rm sw}=1/\tau_{\rm sw}$, which decreases from $6.16$ as $\Pran\to0$ to $5.26$ as $\Pran\to\infty$. The disparity of the molecular diffusivities of heat and light elements in planetary cores motivates this unequal-diffusivity regime, whereas codensity models assume that turbulent mixing equalises the two diffusivities; the critical Lewis numbers refer to the annulus.

\begin{figure}[!htbp]
 \centering\includegraphics[width=\textwidth]{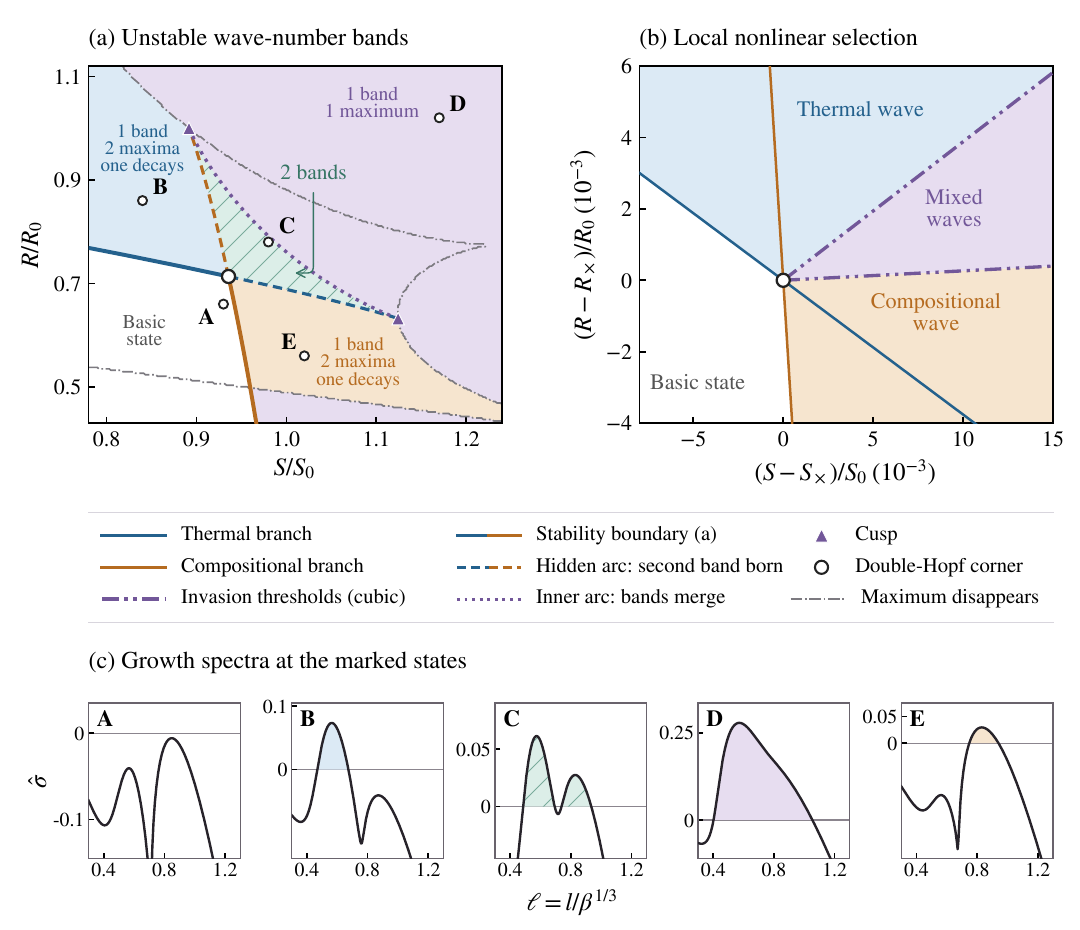}
 \caption{Spectral geometry and nonlinear selection at $\Pran=0.3$, $\tau=0.1$, $\beta=3.786\times10^6$, $k=\pi$, normalised by \eqref{eq:pure}. (a) Finite-rotation spectral map: white is stable; hatching marks two separated unstable bands. (b) Cubic nonlinear selection near the reference corner: the purple dash-dotted rays delimit the cubic coexistence wedge, inside which both waves coexist as a stable mixed state; white is the stable basic state. (c) Spectra $\hat\sigma=\sigma/\beta^{2/3}$ at A--E in (a); growth is shaded; vertical scales differ. A--D pass from stability to merged bands; E is the compositional band. Plotted with the AI-assisted scripts described in Appendix~\ref{sec:numerics}.}
 \label{fig:selection}
\end{figure}

\section{Reading the spectrum from the front}\label{sec:front}
The exposed part of the envelope is the onset threshold. The hidden part is invisible at onset, but it is not idle: it records how the set of unstable wave numbers changes as the forcing grows. This section reads the whole front as a map of the unstable spectrum and then examines the compositional end of the exposed boundary, where heat can stabilise.

Let $\mathcal U=\{l:\sigma>0\}$ be the set of unstable wave numbers, where $\sigma(l;R,S)$ is the largest growth rate defined in Sec.~\ref{sec:envelope}. Subscripts denote partial derivatives at fixed $\Pran$, $\tau$, $\beta$ and $k$, so that $\sigma_{lR}=\partial^2\sigma/\partial l\,\partial R$.

\begin{proposition}[spectral strata]\label{prop:strata}
Suppose the leading root is unique, simple and smooth, and $\sigma_S\sigma_{lR}-\sigma_R\sigma_{lS}\neq0$ on the envelope. Away from the endpoints of the wave-number interval, the number of components of $\mathcal U$ changes only on $\sigma=\sigma_l=0$. When the controls move towards positive growth at the tangency, a neutral maximum, $\sigma_{ll}<0$, creates an unstable band, and a neutral minimum, $\sigma_{ll}>0$, merges two bands. An ordinary cusp has $\sigma_{ll}=0$ and $\sigma_{lll}\neq0$, and a self-intersection has two distinct neutral wave numbers.
\end{proposition}

Differentiating $\sigma=\sigma_l=0$ along the envelope $(R(l),S(l))$ gives its tangent,
\begin{equation}
 \left(\frac{\rd R}{\rd l},\frac{\rd S}{\rd l}\right)=-\frac{\sigma_{ll}}{\sigma_S\sigma_{lR}-\sigma_R\sigma_{lS}}
                      (\sigma_S,-\sigma_R).
 \label{eq:frontvelocity}
\end{equation}
The tangent vanishes where $\sigma_{ll}=0$; with $\sigma_{lll}\neq0$ this is an ordinary cusp. The proposition turns the front into a map of the spectrum, shown in Fig.~\ref{fig:selection}(a) at finite rotation. Crossing an exposed arc towards positive growth creates the first unstable band at a neutral maximum. Crossing a hidden outer arc creates a second band, centred on another wave number, while the first band already grows. Crossing the hidden inner arc merges the two bands at a neutral minimum. Two maxima may survive the merger, until one of them disappears at the dash-dotted locus. Figure~\ref{fig:selection}(c) shows the growth spectra at the marked states. With discrete wave numbers $l=m\Delta l$, onset is selected among the admissible $l$ only, so the neutrality of another family, hidden on the continuous front, can become the observed threshold when no admissible $l$ lies in the band that grows first.

\subsection{When heat stabilises the compositional wave}\label{sec:tangent}
At its compositional end the exposed boundary of the leading rapid-rotation front has the explicit tangent
\begin{equation}
 \left.\frac{\rd S_{\rm crit}}{\rd R}\right|_{R=0}
 =\frac{1-4a\tau/b}{1+2\tau^2},\qquad
 \tau_*=\frac{\Pran}{4\Pran+3}.
 \label{eq:tangent}
\end{equation}
The slope changes sign at $\tau_*$. For $\tau<\tau_*$, adding destabilising heat raises the compositional threshold: heat stabilises the compositional wave. The reason is the phase of the thermal response. At a simple root of \eqref{eq:cubic} the growth responds to $R$ as
\begin{equation*}
\partial_R\lambda=\frac{l^2\,T_T}{\mathcal F_\lambda},\qquad \mathcal F_\lambda=\frac{\partial\mathcal F}{\partial\lambda},
\end{equation*}
so heat adds growth only through the part of its response that is in phase with $\mathcal F_\lambda$. At the compositional onset $X=-\sqrt2\,\tau$, and the thermal and compositional responses have phases $\arctan(\sqrt2\,\tau)$ and $\arctan\sqrt2$. Heat reduces the growth when its phase differs from that of $\mathcal F_\lambda$ by more than $\pi/2$, and this is exactly the condition $\tau<\tau_*$ in \eqref{eq:tangent}. Section~\ref{sec:sphere} compares the slope with spherical calculations.

\section{From the kink to coexistence}\label{sec:nonlinear}
At a corner two waves become neutral together, and just above it both can grow. Do they compete, so that one suppresses the other, or do they coexist? This section shows that the answer is already contained in the linear geometry of the corner. The interaction through the mean scalar profiles factorises into the normals of the two neutral arcs and the responses of the two scalars, and the resulting coexistence margin is positive at every corner of the leading front.

\subsection{Two waves and their amplitude equations}\label{sec:amplitudes}
Near a corner, the stream function is dominated by the two critical waves, labelled $j,k\in\{T,C\}$,
\begin{equation*}
\psi=\sum_{j=T,C}A_j\sin(\pi x)\exp[\mathrm{i}(l_jy+\omega_jt)]+\mathrm{c.c.},
\end{equation*}
with slowly varying complex amplitudes $A_T$ and $A_C$. At the reference corner their wave-number ratio $\ell_C/\ell_T=1.486$ avoids the ratios $1$, $2$ and $3$ that would admit resonant quadratic or cubic terms. Absorbing the linear detunings into the phases then gives the amplitude equations
\begin{equation}
 \dot A_j=A_j\left(\mu_j+\sum_{k=T,C}c_{jk}|A_k|^2\right),
 \qquad j=T,C,
 \label{eq:amp}
\end{equation}
the non-resonant double-Hopf normal form \citep[\S~8.6]{Kuznetsov2023}. Here $\mu_T$ and $\mu_C$ are the real growth rates of the two waves at their fixed corner wave numbers $l_T$ and $l_C$; both vanish at the corner $(R_\times,S_\times)$. The self coefficients $c_{TT}$ and $c_{CC}$ measure how each wave saturates, and the cross coefficients $c_{TC}$ and $c_{CT}$ how each wave changes the growth rate and frequency of the other. \citet[\S~4]{Busse1986} derived the analogue of \eqref{eq:amp} for radial modes $n=1,2$ of one scalar (see also Ref.~\cite{Sanchez2013}); our waves instead differ in $l$, and they couple through the mean scalar profiles rather than through a mean zonal flow.

With $I_j=|A_j|^2$ and $g_{jk}=-\mathrm{Re}\,c_{jk}$, the real parts of \eqref{eq:amp} give the intensity equations
\begin{equation}
 \dot I_j=2I_j\Bigl(\mu_j-\sum_{k=T,C}g_{jk}I_k\Bigr),\qquad j=T,C.
 \label{eq:intensity}
\end{equation}
When all $g_{jk}>0$ and $\mu_T>0$, the mixed state exists and is stable exactly when
\begin{equation}
 \chi_{\rm full}=\frac{g_{TC}g_{CT}}{g_{TT}g_{CC}}<1,
 \qquad
 \frac{g_{CT}}{g_{TT}}<\frac{\mu_C}{\mu_T}<\frac{g_{CC}}{g_{TC}}.
 \label{eq:wedge}
\end{equation}
The inequalities make both mixed intensities $(I_T,I_C)^{\mathsf T}=G_{\rm full}^{-1}(\mu_T,\mu_C)^{\mathsf T}$ positive, with $G_{\rm full}=(g_{jk})$, and the Jacobian $-2\operatorname{diag}(I_T,I_C)G_{\rm full}$ then has negative trace and positive determinant. The two edges of the wedge are invasion thresholds: a small wave $j$ invades the saturated pure wave $k$, of intensity $\mu_k/g_{kk}$, at the rate $\mu_j-g_{jk}\mu_k/g_{kk}$ \citep{Kuznetsov2023}. In the classification of \citet[\S~8.6.2]{Kuznetsov2023}, with $p_{jk}=-g_{jk}$, the wedge is case II of the simple Hopf--Hopf interaction, and $\chi_{\rm full}>1$ would be the bistable case I. The intensity equations \eqref{eq:intensity} have no periodic orbit in $I_T,I_C>0$, because the Dulac multiplier $1/(I_TI_C)$ gives
\begin{equation}
 \nabla\!\cdot\left(\frac{\dot{\boldsymbol I}}{I_TI_C}\right)
 =-\frac{2g_{TT}}{I_C}-\frac{2g_{CC}}{I_T}<0.
 \label{eq:dulac-intensities}
\end{equation}
A stable mixed state therefore has steady intensities, while the phases of the two waves keep drifting: with unequal phase speeds the pattern is modulated and generically quasi-periodic.

\subsection{Coupling through the mean scalar profiles}\label{sec:means}
Each wave saturates mainly by flattening the interior mean gradients of temperature and concentration. With fixed wall values, the transport by the wave redistributes the gradients and reduces the buoyancy available for growth. Quadratic interaction of each wave with its conjugate in \eqref{eq:nl} creates these mean distortions. Vorticity self-advection vanishes, because $\nabla^2\psi_k=-H_k\psi_k$ with $H_k=\pi^2+l_k^2$, and so do the second azimuthal harmonics. This separation of mean-field and harmonic feedback, and saturation by mean-gradient flattening, follow weakly nonlinear analyses of spherical shells \citep{Skene2025}. Phase averaging gives $\bar\Theta_k''=\Pran\,\langle J(\psi_k,\Theta_k)\rangle$ and $\tau\bar\Xi_k''=\Pran\,\langle J(\psi_k,\Xi_k)\rangle$. With the responses $T_{T,k}=(\mathrm{i}\Pran\,\omega_k+H_k)^{-1}$ and $T_{C,k}=(\mathrm{i}\Pran\,\omega_k+\tau H_k)^{-1}$ of the wave $k$, whose real parts are $H_k|T_{T,k}|^2$ and $\tau H_k|T_{C,k}|^2$, integration with zero wall values gives the mean profiles
\begin{equation}
 \bar\Theta_k=-\frac{\Pran\,l_k^2H_k|T_{T,k}|^2}{2\pi}|A_k|^2\sin2\pi x,
 \qquad
 \bar\Xi_k=-\frac{\Pran\,l_k^2H_k|T_{C,k}|^2}{2\pi}|A_k|^2\sin2\pi x.
 \label{eq:means}
\end{equation}
Advection of these profiles by wave $j$, projected onto the left null vector of \eqref{eq:M}, gives how the flattening by wave $k$ changes the growth rate and frequency of wave $j$:
\begin{equation}
 \begin{aligned}
 c^{(0)}_{jk}&=-\frac{\Pran^2\,l_j^2l_k^2H_k}{2}
 \frac{R T_{T,j}|T_{T,k}|^2+\tau S T_{C,j}|T_{C,k}|^2}
 {\mathcal F_{\lambda,j}},\\
 \mathcal F_{\lambda,j}&=H_j+\Pran\,l_j^2(RT_{T,j}^2+\tau ST_{C,j}^2).
 \end{aligned}
 \label{eq:cjk}
\end{equation}
The superscript marks the zero wave number of the means. Because the second harmonics vanish, \eqref{eq:cjk} with $j=k$ gives the self coefficients exactly, and they saturate.

\begin{proposition}[positive self-saturation]\label{prop:self}
For every neutral wave with $l>0$ in $R,S>0$, at any rotation rate, the self coefficient $g_{jj}=-\mathrm{Re}\,c_{jj}$ is positive.
\end{proposition}

\emph{Proof.} For a neutral wave put $T_T=(H+\mathrm{i}\Pran\,\omega)^{-1}$, $T_C=(\tau H+\mathrm{i}\Pran\,\omega)^{-1}$,
\begin{equation}
 M_2=R|T_T|^2+\tau S|T_C|^2,\qquad M_4=R|T_T|^4+\tau^2S|T_C|^4,
 \label{eq:M2-M4}
\end{equation}
$B_4=R|T_T|^4+\tau S|T_C|^4$ and $D_4=R|T_T|^4+\tau^3S|T_C|^4$. The numerator of \eqref{eq:cjk} with $j=k$ is $HM_4-\mathrm{i}\Pran\,\omega B_4$, while $\mathrm{Re}\,\mathcal F_\lambda=H+\Pran\,l^2(H^2D_4-\Pran^2\omega^2B_4)$ and $\mathrm{Im}\,\mathcal F_\lambda=-2\Pran^2\omega\,l^2HM_4$. Hence
\begin{equation*}
 \mathrm{Re}\,\frac{HM_4-\mathrm{i}\Pran\,\omega B_4}{\mathcal F_\lambda}
 =\frac{HM_4}{|\mathcal F_\lambda|^2}\Bigl[H+\Pran\,l^2\bigl(H^2D_4+\Pran^2\omega^2B_4\bigr)\Bigr]
 =\frac{HM_4}{|\mathcal F_\lambda|^2}\bigl(H+\Pran\,l^2M_2\bigr),
\end{equation*}
where the last step uses $(H^2+\Pran^2\omega^2)|T_T|^4=|T_T|^2$ and $(\tau^2H^2+\Pran^2\omega^2)|T_C|^4=|T_C|^2$. Therefore
\begin{equation}
 g_{jj}=\frac{\Pran^2l^4H^2}{2|\mathcal F_\lambda|^2}\,M_4\,\bigl(H+\Pran\,l^2M_2\bigr)>0,
 \label{eq:gjj-positive}
\end{equation}
since every factor is positive and $\mathcal F_\lambda\neq0$ by \eqref{eq:neutral-root-simplicity}.\hfill$\square$

For the cross terms, harmonics at $(l_T\pm l_C,\omega_T\pm\omega_C)$ add further contributions to \eqref{eq:cjk}; Sec.~\ref{sec:finite} evaluates them.

\subsection{The geometric criterion}\label{sec:criterion}
The mean-profile coefficients \eqref{eq:cjk} are built from the same responses that set the linear growth. Near the corner the growth rates at fixed corner wave numbers depend linearly on the forcing through the Jacobian $J_\mu$ at the corner,
\begin{equation}
 \boldsymbol\mu=J_\mu\begin{pmatrix}R-R_\times\\S-S_\times\end{pmatrix}
       +O(\|(R-R_\times,S-S_\times)\|^2),\qquad
 (J_\mu)_{jR}=l_j^2\operatorname{Re}\frac{T_{T,j}}{\mathcal F_{\lambda,j}},\quad
 (J_\mu)_{jS}=\tau l_j^2\operatorname{Re}\frac{T_{C,j}}{\mathcal F_{\lambda,j}}.
 \label{eq:mu-sensitivities}
\end{equation}
The rows $\boldsymbol n_T$ and $\boldsymbol n_C$ of $J_\mu$ are the normals to the two neutral arcs at the corner. Both waves are stationary in $l$ there, so these are also the normals to the front, and $\det J_\mu=\|\boldsymbol n_T\|\,\|\boldsymbol n_C\|\sin\theta_{\rm kink}$ measures the transverse crossing of the arcs through the oriented kink angle $\theta_{\rm kink}$ from $\boldsymbol n_T$ to $\boldsymbol n_C$. Normalising by $R_0$ and $S_0$ changes the numerical value of $\theta_{\rm kink}$ but neither the sign of $\det J_\mu$ nor the order at which it closes. The map \eqref{eq:mu-sensitivities} sends the edges of the wedge \eqref{eq:wedge} to the purple rays of Fig.~\ref{fig:selection}(b).

Since $|T_{T,k}|^2=[H_k^2(1+X_k^2)]^{-1}$ and $|T_{C,k}|^2=[H_k^2(\tau^2+X_k^2)]^{-1}$ are real, taking real parts of \eqref{eq:cjk} factors the mean-profile saturation matrix $G^{(0)}=(g^{(0)}_{jk})$, with $g^{(0)}_{jk}=-\mathrm{Re}\,c^{(0)}_{jk}$, through the growth sensitivities:
\begin{gather}
 G^{(0)}=\frac{\Pran^2}{2}J_\mu\operatorname{diag}(R,S)\mathcal W,
 \qquad
 \mathcal W=\begin{pmatrix}
 l_T^2H_T|T_{T,T}|^2&l_C^2H_C|T_{T,C}|^2\\
 l_T^2H_T|T_{C,T}|^2&l_C^2H_C|T_{C,C}|^2
 \end{pmatrix},
 \label{eq:factorG}\\
 \det G^{(0)}=\frac{\Pran^4}{4}RS\det J_\mu\det\mathcal W,
 \label{eq:detG}\\
 \det\mathcal W=\frac{l_T^2l_C^2(1-\tau^2)(X_T^2-X_C^2)}
 {H_TH_C(1+X_T^2)(1+X_C^2)(\tau^2+X_T^2)(\tau^2+X_C^2)}.
 \label{eq:detW}
\end{gather}
The determinant \eqref{eq:detG} separates the crossing of the two neutral arcs, through $\det J_\mu$, from the contrast in scalar feedback, through $\det\mathcal W$: the difference between the waves in how strongly they flatten each scalar. For $0<\tau<1$ the second factor vanishes only when $X_T^2=X_C^2$, that is, when the two waves flatten the two scalars in the same proportion. In the nonrotating Veronis model, eliminating the mean temperature and concentration modes turns a quantity proportional to the kinetic energy of a steady roll into an eigenvalue of a matrix related to the linear one \citep{Tuckerman2001,Tuckerman2012}. Here two waves distort the scalar profiles in different proportions: $\mathcal W$ records how strongly each wave flattens each scalar, and $J_\mu$ how the growth of each wave responds. If both waves saturated through a single common channel, $G^{(0)}$ would have rank one and there would be no positive coexistence margin.

\begin{proposition}[frequency parameters]\label{prop:frequency}
Let $R,S>0$, $0<\tau<1$ and $\beta>0$. At fixed $\Pran$, $\tau$, $R$, $S$ and $\beta$, a neutral wave with $l>0$ is determined by its frequency parameter $X<0$. Two distinct neutral waves therefore have $X_T^2\neq X_C^2$ and $\det\mathcal W\neq0$. In the leading rapid-rotation problem the broad thermal wave has the larger $X^2$, so $\det\mathcal W>0$.
\end{proposition}

\emph{Proof.} With $\mathcal A(X)=R/(1+X^2)+\tau^2S/(\tau^2+X^2)$ and $\mathcal B(X)=R/(1+X^2)+\tau S/(\tau^2+X^2)$, the real and imaginary parts of $\mathcal F=0$ read $\mathcal A=H^3/l^2$ and $\mathcal B=-\mathcal A/\Pran-\beta H/(lX)$ (Appendix~\ref{sec:quartic}). Hence $q=H/l=-(X/\beta)(\mathcal B+\mathcal A/\Pran)$, and then $H=\mathcal A/q^2$ and $l=\mathcal A/q^3$: the wave is fixed by $X$. In the leading problem the real condition becomes $\ell^4=\hat R/(1+X^2)+\tau^2\hat S/(\tau^2+X^2)$, which decreases with $X^2$, so $\ell_T<\ell_C$ gives $X_T^2>X_C^2$.\hfill$\square$

The two factors combine into one coexistence margin. Let $\chi^{(0)}=g^{(0)}_{TC}g^{(0)}_{CT}/(g^{(0)}_{TT}g^{(0)}_{CC})$ and $Z_j=X_j^2$, and define for each wave the weighted sensitivity ratio and the scalar-response ratio,
\begin{equation*}
\rho_j=\frac{S\,\partial_S\mu_j}{R\,\partial_R\mu_j},\qquad
\gamma_j=\frac{|T_{C,j}|^2}{|T_{T,j}|^2}=\frac{1+Z_j}{\tau^2+Z_j}.
\end{equation*}
Removing row and column factors from \eqref{eq:factorG} gives
\begin{equation}
 \mathcal M^{(0)}\equiv1-\chi^{(0)}
 =\frac{(\rho_C-\rho_T)(\gamma_C-\gamma_T)}
 {(1+\rho_T\gamma_T)(1+\rho_C\gamma_C)}.
 \label{eq:margin-ratios-main}
\end{equation}
The first difference compares how the two waves draw on the two buoyancies, and the second compares how strongly each wave flattens composition relative to temperature. When both waves draw positive growth from heat, coexistence requires the two differences to have the same sign: the wave that relies more on composition must also flatten composition more. It requires $\partial_R\mu_T\,\partial_R\mu_C\neq0$; the homogeneous form \eqref{eq:margin-homogeneous} below removes this restriction. The identity holds at any rotation rate.

\begin{figure}[!htbp]
\centering\includegraphics[width=.8\textwidth]{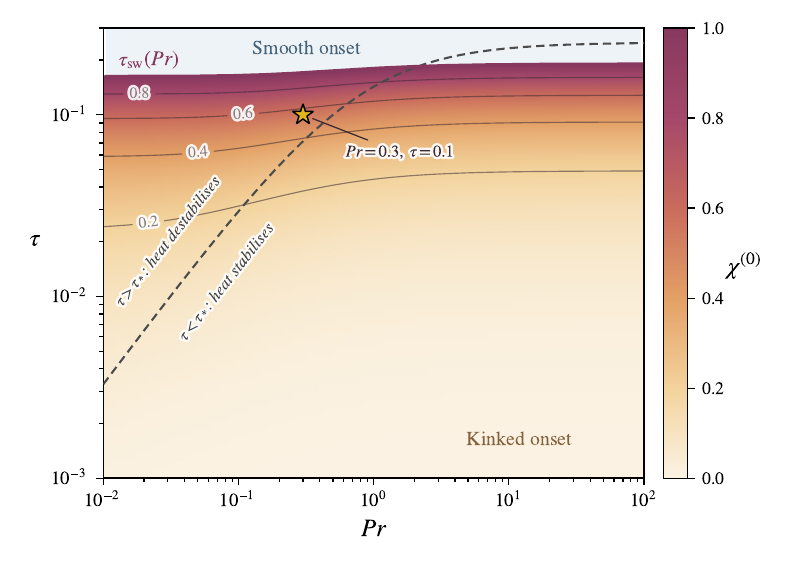}
\caption{Mean-profile coupling ratio $\chi^{(0)}=1-\mathcal M^{(0)}$ at the leading corners, from \eqref{eq:margin-exact}, with labelled contours. Above the swallowtail locus $\tau_{\rm sw}(\Pran)$ of Fig.~\ref{fig:sw-locus} the onset is smooth and has no corner. Equation~\eqref{eq:margin-exact} gives $\chi^{(0)}<1$ for distinct waves, so in the leading problem the mean profiles favour coexistence at every corner of the front; $\chi^{(0)}\to1$ only at the numerically continued locus, where the two waves merge. Section~\ref{sec:merger} treats the complete cubic interaction near merger. The dashed curve $\tau_*=\Pran/(4\Pran+3)$ marks the sign change of the compositional endpoint slope in equation~\eqref{eq:tangent}: below it added heat raises the compositional threshold, above it heat lowers it. Crossing $\tau_*$ changes the compositional-end sensitivity without changing the mean-profile coexistence criterion at the resolved corners. The star is $\Pran=0.3$, $\tau=0.1$. Plotted with the AI-assisted scripts described in Appendix~\ref{sec:numerics}.}
\label{fig:map}
\end{figure}

\begin{proposition}[positive margin]\label{prop:margin}
At every corner of the leading rapid-rotation front in $R,S>0$, $0<\tau<1$,
\begin{equation}
 \mathcal M^{(0)}
 =\frac{3(2\tau^2+Z_TZ_C)(Z_T-Z_C)^2}
 {4(Z_T+\tau^2)(Z_C+\tau^2)(1+Z_T)(1+Z_C)}>0.
 \label{eq:margin-exact}
\end{equation}
Hence $\det G^{(0)}>0$, and with the waves ordered by $\ell_T<\ell_C$, $\det\mathcal W>0$ and $\det J_\mu>0$.
\end{proposition}

\emph{Proof.} Set $u_j=R\,\partial_R\mu_j$ and $v_j=S\,\partial_S\mu_j$. Writing $\mathcal W_{2j}=\gamma_j\mathcal W_{1j}$ in \eqref{eq:factorG}, the row and column factors cancel from $\chi^{(0)}$:
\begin{equation}
 1-\chi^{(0)}=
 \frac{(u_Tv_C-v_Tu_C)(\gamma_C-\gamma_T)}
 {(u_T+v_T\gamma_T)(u_C+v_C\gamma_C)}.
 \label{eq:margin-homogeneous}
\end{equation}
Each factor of the denominator is a positive multiple of a self coefficient, which is positive by Proposition~\ref{prop:self}, so \eqref{eq:margin-homogeneous} stays regular where a single sensitivity vanishes; dividing by $u_Tu_C$ gives \eqref{eq:margin-ratios-main}. Write the two-parameter family \eqref{eq:lead} in the coordinates $(w,Z)$, before the radical \eqref{eq:radicalZ} is substituted. On the envelope $\sigma_\ell=0$, so the normal annihilates every variation of this family, in particular its $Z$-variation at fixed $w$; the derivatives below are taken in the family and then restricted to the envelope. Since
\begin{equation}
 \partial_Z\log\hat R=\frac{Z-2}{3Z(1+Z)},\qquad
 \partial_Z\log\hat S=\frac{Z-2\tau^2}{3Z(Z+\tau^2)},
 \label{eq:log-family-normal}
\end{equation}
the sensitivity ratio is
\begin{equation}
 \rho(Z)=\frac{(2-Z)(Z+\tau^2)}{(1+Z)(Z-2\tau^2)}.
 \label{eq:rho-leading}
\end{equation}
Substitution into \eqref{eq:margin-ratios-main} gives \eqref{eq:margin-exact}; at $Z=2\tau^2$, where $\rho$ has a pole, its value follows by continuity from \eqref{eq:margin-homogeneous}. Its numerator and denominator are positive because $Z_T\neq Z_C$ by Proposition~\ref{prop:frequency}. With positive self coefficients this gives $\det G^{(0)}>0$, and Proposition~\ref{prop:frequency} with \eqref{eq:detG} gives the orientation.\hfill$\square$

The proposition answers the question of this section at leading order: wherever the threshold has a corner, the mean profiles favour coexistence. Because $\det J_\mu\neq0$, positive mixed intensities $(I_T,I_C)^{\mathsf T}=(G^{(0)})^{-1}\boldsymbol\mu$ occur in an open region of controls near the corner, and there the mixed state is stable in the mean-profile cubic equations: its Jacobian $-2\operatorname{diag}(I_T,I_C)G^{(0)}$ has negative trace and positive determinant. No assumption on the connectedness of the set of corners is needed. The individual sensitivities need not be positive: at $\Pran=0.3$ and $\tau=0.03$, for instance, added heat reduces the growth rate of the compositional wave, $\rho_C<0$, and the margin is still $0.843$. Figure~\ref{fig:map} illustrates \eqref{eq:margin-exact} over the kinked part of the $(\Pran,\tau)$ plane. The margin is large deep in the kinked regime and closes only at the swallowtail locus, where the two waves merge.

\section{Near the swallowtail}\label{sec:finite}
The margin \eqref{eq:margin-exact} is positive at every corner, but it vanishes as the two waves merge at the swallowtail. Near the merger the harmonic feedback, which the mean-profile coefficients leave out, can decide the interaction. This section derives how the margin closes and evaluates the complete cubic coefficients at finite rotation.

\subsection{The closing margin}\label{sec:closing}
By the normal form of Sec.~\ref{sec:normal-form}, the wave-number separation and the kink angle scale as $(\tau_{\rm sw}-\tau)^{1/2}$ near the swallowtail. On the leading front the real neutral condition $\ell^4=\hat R/(1+Z)+\tau^2\hat S/(\tau^2+Z)$ decreases strictly with $Z$, so $Z_T-Z_C$ is of first order in the wave-number separation. Equation \eqref{eq:margin-exact} is quadratic in $Z_T-Z_C$, and therefore in that angle, so the margin closes linearly in the distance from the swallowtail,
\begin{equation}
 \mathcal M^{(0)}\sim C_{\rm m}^{\rm lead}(\tau_{\rm sw}-\tau),\qquad \tau\uparrow\tau_{\rm sw},
 \label{eq:closing}
\end{equation}
with an algebraic coefficient, $C_{\rm m}^{\rm lead}=6.98827$ at $\Pran=0.3$ (Appendix~\ref{sec:margin-coefficient}). At $\beta=3.786\times10^6$ the finite-rotation law has $\tau_c-\tau$ in place of $\tau_{\rm sw}-\tau$ and the coefficient $C_{\rm m}(\Pran,\beta)=6.98941$. The margin is thus small only in a thin layer below $\tau_{\rm sw}$, and there the small harmonic contributions can decide the interaction.

\subsection{The complete cubic coefficients}\label{sec:projection}
For distinct waves, the cubic forcing has three parts: the mean profiles \eqref{eq:means}, and the difference and sum harmonics at $(l_T\mp l_C,\omega_T\mp\omega_C)$. All three have radial index two, and their products with a critical wave contain radial indices one and three only, so the cubic coefficients close without radial truncation. Projecting the resonant forcing onto the left null vector gives
\begin{equation}
 c_{jk}=c^{(0)}_{jk}+c^-_{jk}+c^+_{jk},\qquad j\neq k,
 \label{eq:full-coefficient}
\end{equation}
where each term is a projection through a $3\times3$ modal inverse (Appendix~\ref{app:cubic}). The self coefficients have no harmonic part, so they remain those of Proposition~\ref{prop:self}.

\begin{table}[!htbp]
\centering
\caption{Mean-profile and complete cubic coupling at $\Pran=0.3$, $\beta=3.786\times10^6$. Coefficients and all unrounded corner coordinates are available from the author on request.}\label{tab:interaction}
\begin{tabular}{ccccc}
\toprule
$\tau$ & $l_T$ & $l_C$ & $\chi^{(0)}$ & $\chi_{\rm full}$\\
\midrule
0.100000000 & 88.788393 & 131.932697 & 0.54988945 & 0.55037629\\
0.169000000 & 102.429884 & 107.439029 & 0.99269809 & 0.99230068\\
0.169546187 & 103.166362 & 106.627966 & 0.99650765 & 0.99663283\\
0.169846187 & 103.782289 & 105.971139 & 0.99860250 & 1.00024294\\
0.169946187 & 104.096081 & 105.643723 & 0.99930115 & 1.00342286\\
\bottomrule
\end{tabular}
\end{table}

\begin{figure}[!htbp]
\centering\includegraphics[width=\textwidth]{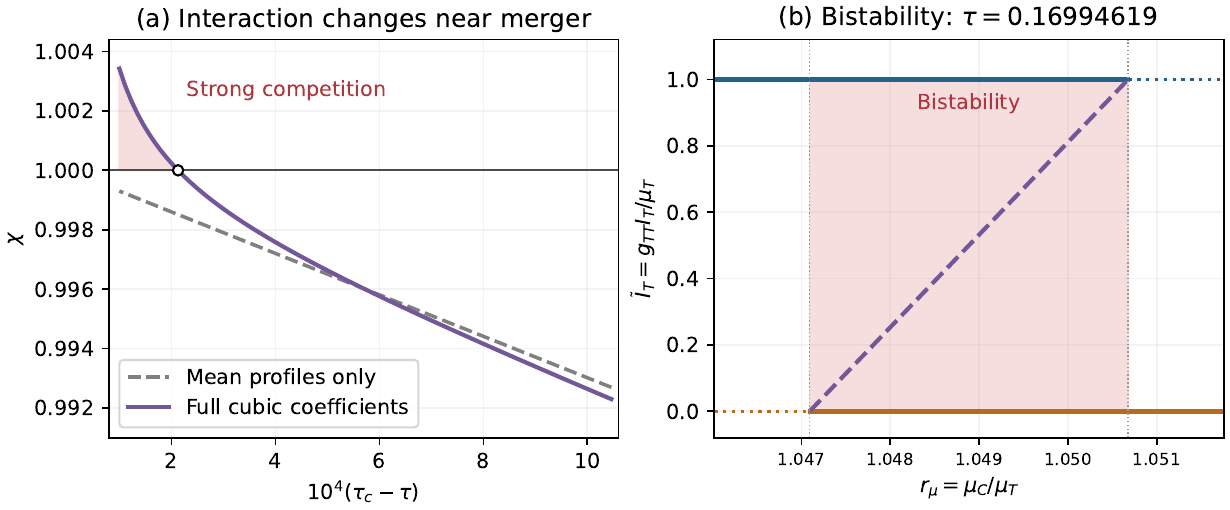}
\caption{Prediction of the cubic two-mode equations near the finite-rotation merger, $\Pran=0.3$, $\beta=3.786\times10^6$. (a) Full and mean-profile coupling ratios against the distance from the merger $\tau_c=0.1700462$. (b) Normalised thermal intensity $\widetilde I_T=g_{TT}I_T/\mu_T$ at $\tau=0.16994619$, where the stable pure branches overlap: solid pure-wave branches are stable, dotted continuations are unstable, and the dashed purple mixed branch is a saddle. This narrow bistable interval has not been tested by the spatial simulations. Plotted with the AI-assisted scripts described in Appendix~\ref{sec:numerics}.}
\label{fig:bistability}
\end{figure}

At the reference corner, given by the annulus parameters of \citet{Monville2019} ($\Pran=0.3$, $\Sch=3$, $\beta=3.786\times10^6$), the complete saturation matrix and the coexistence wedge are
\begin{equation}
 G_{\rm full}=\begin{pmatrix}
 547.960&8148.23\\
 504.962&13643.1
 \end{pmatrix},\qquad
 0.92153<\frac{\mu_C}{\mu_T}<1.67436,
 \label{eq:reference-wedge}
\end{equation}
with $\chi_{\rm full}=0.55038$ against $\chi^{(0)}=0.54989$ from the mean profiles. The harmonics change the two cross coefficients by $-0.181\%$ and $+0.270\%$. Since the self coefficients are exact, any relative change $\varepsilon\leq0.003$ of the cross coefficients gives $\chi_{\rm full}\leq\chi^{(0)}(1+\varepsilon)^2<0.554$, so the harmonics cannot overturn coexistence there. Figure~\ref{fig:selection}(b) shows this wedge at the reference corner.

Mean fields alone need not suffice: in non-rotating thermosolutal convection with opposing gradients, linearization about the mean fields predicts the frequencies of travelling waves but not of standing waves \citep{Turton2015}, and a self-consistent model built on the self-interaction of the leading mode reproduces the frequency accurately near onset but not across the travelling-wave branch \citep{Bengana2021}. In the swirling jet studied by \citet{Meliga2012}, the selection between single and double helices is decided by a competition between harmonics rather than by the mean-flow correction. Here the harmonics are small at the reference corner but can change the prediction near merger (Sec.~\ref{sec:merger}). In spherical shells, zonal flows can make the onset subcritical and a self coefficient negative \citep{Skene2025}; for the pair considered here the quadratic interaction generates no zonal velocity, and Sec.~\ref{sec:sphere} returns to this difference.

The outcome differs from single-scalar systems near mode crossings, which show double columns or bistability in the annulus \citep{Pino2001} and bistable thermal waves with unstable mixed tori in a shell \citep{Sanchez2013}. Here two waves with different frequencies weight the flattening of the two scalars differently; at the reference corner this keeps $g_{TC}g_{CT}<g_{TT}g_{CC}$ with the harmonics included, and coexistence means one stable mixed state.

\subsection{Bistability near merger}\label{sec:merger}
Close to the merger the mean-profile margin is small, and the harmonics become decisive (Table~\ref{tab:interaction}). At $\beta=3.786\times10^6$ the complete interaction reaches $\chi_{\rm full}=1$ at $\tau=0.1698333$, before the merger at finite rotation, $\tau_c=0.1700462$; we write $\tau_c(\Pran,\beta)$ for this merger and keep $\tau_{\rm sw}(\Pran)$ for its rapid-rotation limit. Between the two values the interaction is case I of \citet[\S~8.6.2]{Kuznetsov2023}: the mixed branch is a saddle and the stable pure branches overlap (Fig.~\ref{fig:bistability}). At the point shown in Fig.~\ref{fig:bistability}(b), the wave numbers differ by $1.5\%$ and $\chi_{\rm full}-1\simeq0.0034$; near the first crossing, at $\tau=0.169846$, they differ by $2.1\%$. This is a prediction of the cubic two-mode equations. Fifth-order feedback, neighbouring sidebands and the admissible wave-number spacing can alter it, and the simulations of Sec.~\ref{sec:validation} test the robust coexistence wedge at $\tau=0.1$ instead.

For an isolated non-resonant pair with a regular unfolding, non-zero cubic coefficients $g_{jk}$ and $\det G_{\rm full}\neq0$, the Hopf and secondary bifurcation curves of the cubic equations persist locally with the same tangents \citep[\S~8.6.3, lemma~8.16]{Kuznetsov2023}. The statement is local and assumes the spectral gap of an isolated pair; it is not uniform near $\det G_{\rm full}=0$, which here occurs just before the merger.

\section{Two waves in a periodic cell}\label{sec:validation}
The predictions of Secs.~\ref{sec:nonlinear} and~\ref{sec:finite} concern two waves with continuous wave numbers. A simulation needs a domain in which a definite pair of waves becomes neutral together. We therefore use a periodic cell whose admissible wave numbers make one discrete pair critical, derive the cubic predictions for that pair, and test them against integrations of the full equations \eqref{eq:nl}.

\begin{figure}[!htbp]
 \centering\includegraphics[width=\textwidth]{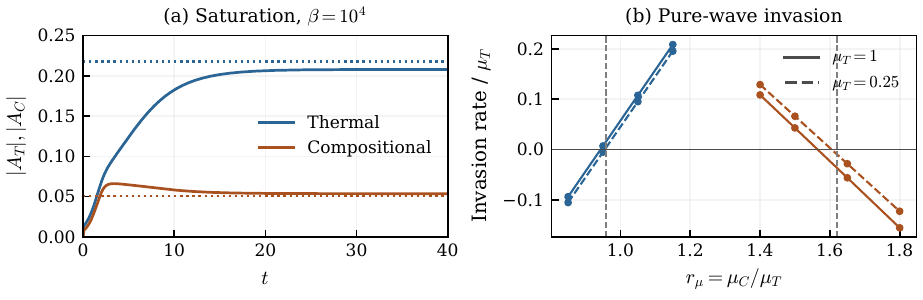}
 \caption{Nonlinear validation near the discrete corner of the periodic cell, $\Pran=0.3$, $\tau=0.1$, $\beta=10^4$, modes $m_T=2$ and $m_C=3$; time and growth use viscous units. (a) Saturation of a mixed state at $(\mu_T,\mu_C)=(1.0058,1.2452)$; dotted lines are cubic amplitudes. (b) Invasion rates about each nonlinear pure wave, coloured by the carrier wave (blue thermal, orange compositional), at $\mu_T=1$ and $0.25$ against $r_\mu=\mu_C/\mu_T$; vertical dashed lines mark the cubic thresholds, which the simulated invasion thresholds approach as $\mu_T$ decreases. Plotted with the AI-assisted scripts described in Appendix~\ref{sec:numerics}.}
 \label{fig:nonlinear}
\end{figure}

\begin{table}[!htbp]
 \centering
\caption{Attracting amplitudes at the discrete corner, $\beta=10^4$. Growth rates are recomputed at each control pair, with cubic coefficients evaluated at the discrete corner \eqref{eq:discrete-wedge}.}\label{tab:dns}
 \begin{tabular}{lccc}
 \toprule
 State & $(\mu_T,\mu_C)$ & simulated $(|A_T|,|A_C|)$ & cubic $(|A_T|,|A_C|)$\\
 \midrule
 Thermal & $(1.0008,0.4801)$ & $(0.286375,4\times10^{-7})$ & $(0.285709,0)$\\
 Compositional & $(0.5045,1.2931)$ & $(7\times10^{-10},0.070626)$ & $(0,0.069120)$\\
 Mixed & $(1.0058,1.2452)$ & $(0.207762,0.053717)$ & $(0.217762,0.050399)$\\
 Weakly mixed & $(0.2500,0.3116)$ & $(0.106049,0.025941)$ & $(0.107350,0.025508)$\\
 \bottomrule
 \end{tabular}
\end{table}

\begin{figure}[!htbp]
\centering\includegraphics[width=.78\textwidth]{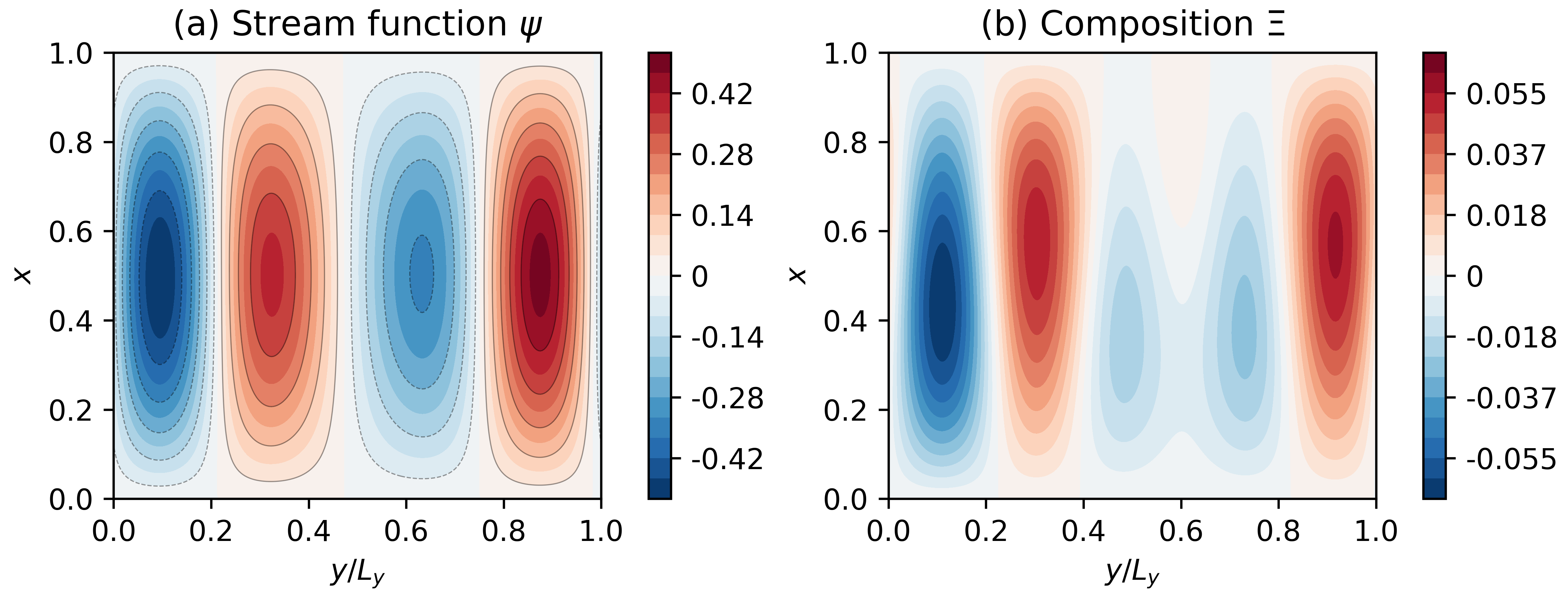}
\caption{The mixed state of Table~\ref{tab:dns}, $(\mu_T,\mu_C)=(1.0058,1.2452)$ at $\beta=10^4$, computed at the refined resolution $(N,M)=(10,10)$ and shown at $t=40$ over one period of the cell: (a) stream function, (b) composition. Both critical waves remain present, and the composition shows the different spatial response of the slower, narrower wave. Plotted with the AI-assisted scripts described in Appendix~\ref{sec:numerics}.}
\label{fig:fields}
\end{figure}

At $\Pran=0.3$, $\tau=0.1$ and $\beta=10^4$, a cell with wave-number spacing $\Delta l=6.11141$ makes the harmonics $m_T=2$ and $m_C=3$, with $l_T=12.2228$ and $l_C=18.3342$, neutral together at $R=4.24885\times10^4$, $S=2.62285\times10^5$, with frequencies $\omega_T=-471.409$ and $\omega_C=-171.518$. All other modes of the cell decay there, the slowest at the rate $3.29$, so the pair is isolated, as the two-mode reduction requires (Appendix~\ref{sec:radial-validity}). This slowest mode is the azimuthally uniform compositional diffusion mode with $n=1$, which decays at $\tau\pi^2/\Pran$. The cubic coefficients evaluated at this discrete corner give
\begin{equation}
 \chi_{\rm full}=0.59209,\qquad
 0.959200<\frac{\mu_C}{\mu_T}<1.620026.
 \label{eq:discrete-wedge}
\end{equation}
No further resonance enters at this order. A phase-dependent term with net amplitude powers $a$ and $b$ in the equation for $A_j$ must match both wave number and frequency, $am_T+bm_C=m_j$ and $a\omega_T+b\omega_C=\omega_j$. Since the phase speeds differ, $m_T\omega_C\neq m_C\omega_T$, and only $(a,b)=(1,0)$ or $(0,1)$ remain. The first additional spatially allowed terms, $\bar A_T^2A_C^2$ and $A_T^3\bar A_C$, would require $3\omega_T=2\omega_C$, whereas here $3\omega_T-2\omega_C=-1071.19$.

\begin{table}[!htbp]
 \centering
\caption{Nonlinear validation beyond Fig.~\ref{fig:nonlinear}. (a) Ratios at which the invasion growth about a Newton-converged pure travelling wave vanishes. They approach the cubic wedge towards onset. (b) Developed mixed states at faster rotation.}\label{tab:validation}
 \begin{tabular}{lccc}
 \toprule
 \multicolumn{4}{l}{(a) Secondary thresholds $\mu_C/\mu_T$ at the discrete corner, $\beta=10^4$}\\
 Invaded wave & $\mu_T=1$ & $\mu_T=0.25$ & cubic limit\\
 \midrule
 Thermal & $0.943002$ & $0.955003$ & $0.959200$\\
 Compositional & $1.565165$ & $1.605274$ & $1.620026$\\
 \midrule
 \multicolumn{4}{l}{(b) Mixed amplitudes divided by cubic values, $(\mu_T,\mu_C)=(2,2.5)$}\\
 $\beta$ & $|A_T|$ & $|A_C|$ & \\
 \midrule
 $10^5$ & $0.918$ & $1.095$ & \\
 $10^6$ & $0.919$ & $1.087$ & \\
 \bottomrule
 \end{tabular}
\end{table}

We integrate \eqref{eq:nl} with a Galerkin expansion in radial sines and azimuthal Fourier modes and fourth-order exponential time differencing (Appendix~\ref{sec:numerics}). The pure waves saturate at the cubic amplitudes to within $0.24\%$ and $2.2\%$. Inside the wedge \eqref{eq:discrete-wedge} both waves grow and saturate together (Fig.~\ref{fig:nonlinear}(a); Table~\ref{tab:dns}). The mixed amplitudes differ from the cubic values by $4.6\%$ and $6.6\%$ at $\mu_T\simeq1$ and by $1.2\%$ and $1.7\%$ at $\mu_T=0.25$, as expected of a truncation whose error is of order $\mu_T$. Figure~\ref{fig:fields} shows the mixed state: both waves are present, and the composition carries the narrower, slower wave more strongly.

The invasion thresholds are tested directly. Newton iteration finds each pure travelling wave in its comoving frame, and the eigenvalues of the full Galerkin Jacobian give the growth of the other wave about it. The zeros of that growth, at $\mu_C/\mu_T=0.9550$ and $1.6053$ for $\mu_T=0.25$, approach the cubic limits \eqref{eq:discrete-wedge} as $\mu_T$ decreases (Fig.~\ref{fig:nonlinear}(b); Table~\ref{tab:validation}). Between them each pure wave is unstable to the other. The cubic equations predict a stable mixed state, and the simulations recover it at the tested controls. Mixed states persist at $\beta=10^5$ and $10^6$ with amplitudes within $10\%$ of the cubic values. Relative to the gap $3.29$, the larger critical growth rate is about $0.095$ in the weakest mixed run but about $0.76$ in the runs at $\beta=10^5$ and $10^6$: the weak runs and the invasion thresholds test the theory near onset, while the faster-rotation runs show that mixed waves persist at the tested controls. The simulations thus confirm the second result where it is robust: two waves of different width and drift coexist as one modulated pattern inside the predicted wedge.

\section{From the annulus to spherical shells}\label{sec:sphere}
The annulus is a model of the equatorial region of a rapidly rotating shell. Its thresholds can be compared with published spherical calculations if each threshold is normalised by its own pure-scalar onsets, and its waves if only scale-free ratios are compared. This section shows what carries over and what does not.

\begin{figure}[!htbp]
\centering\includegraphics[width=.64\textwidth]{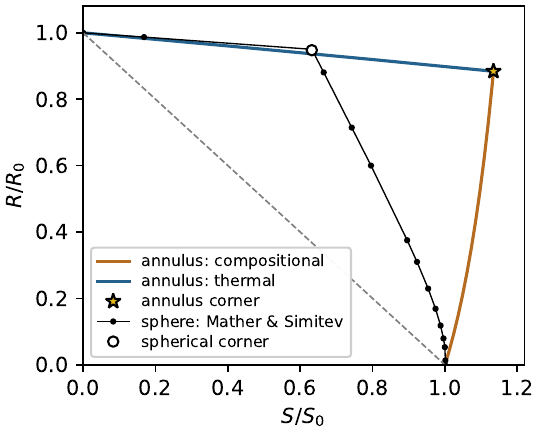}
\caption{Annulus onset and the critical curve of the shell of \citet{Mather2021}, read from their Fig.~1(b), at $\Pran=1$, $\Sch=25$, each normalised by its own pure-scalar thresholds. The star and the open circle mark the annulus and spherical corners, and the dashed line is the additive reference $R/R_0+S/S_0=1$. Both thresholds are non-additive and switch to narrower, slower compositional modes. On the thermal side both arms are shallow and negative: in these coordinates the annulus endpoint tangent has slope $-0.0975$ and the secant of the spherical curve $-0.0795$. Plotted with the AI-assisted scripts described in Appendix~\ref{sec:numerics}.}
\label{fig:sphere}
\end{figure}

\paragraph{Mather and Simitev (2021).} Figure~\ref{fig:sphere} compares the annulus front with the critical curve of the shell of \citet{Mather2021} at $\Pran=1$, $\Sch=25$. Both thresholds are non-additive, with corner sums $2.02$ and $1.58$, and both switch to a narrower, slower compositional mode. At the thermal end, in the normalised coordinates of Fig.~\ref{fig:sphere}, the annulus endpoint tangent has slope $-0.0975$, and the secant of the spherical thermal arc between its vertices read at $R_C\approx0.80\times10^5$ and $2.98\times10^5$ has slope $-0.0795$: both thermal arms are shallow and negative. The scale-free ratios compared below with \citet{Net2012} are the sharper quantitative test. The spherical compositional arc turns inward, and its endpoint slope is not resolved by the published figure. The compositional endpoint slope \eqref{eq:tangent}, with its phase-controlled sign change at $\tau_*$, therefore remains a prediction for shells.

\paragraph{Net et al.\ (2012).} \citet{Net2012} computed onset in a shell with radius ratio $0.35$, $E=10^{-3}$, $\Pran=0.1$ and $\tau=0.01$. Their critical curve over the azimuthal order $m$ consists of two families of neutral curves. At the interchange, $m$ jumps from $3$ to $6$ and the magnitude of the prograde drift frequency falls from $97.24$ to $3.822$, by a factor $25.4$. At the same $\Pran$ and $\tau$, the leading annulus corner has $\ell_C/\ell_T=2.251$ and $\omega_T/\omega_C=19.6$. These scale-free ratios test the direction and the approximate size of the mode switch without fitting a Rayleigh or a frequency scale.

\paragraph{Breuer et al.\ (2010).} The shell simulations of \citet{Breuer2010} use $\Pran=0.3$, $\Sch=3$, rigid walls and $E=10^{-3}$. Their estimate of about $50\%$ more columns for the pure compositional branch than for the pure thermal branch compares with the ratio $\ell_C/\ell_T=1.49$ of the two waves at the annulus corner. Their scans at fixed total Rayleigh number and thermal share $\delta$ are rays $R/S=\Pran\,\delta/[\Sch(1-\delta)]$. With the annulus corner expressed through their pure-scalar thresholds, the ray through it has $\delta^*=0.840$; this calibration makes the number a consistency check rather than an independent prediction. After the same calibration, the annulus thermal-end tangent changes the threshold along their rays by less than $5\%$ for $\delta\geq0.95$, consistent with their simulations at $\delta=0.95$ and $0.99$, whose Nusselt numbers and kinetic energies stay close to purely thermal driving. Their regime change near $\delta=0.8$ belongs to developed flow, away from onset. At the same $\Pran=0.3$ and $\Sch=3$, \citet{Kannan2026} found a change from thermally to compositionally dominated developed convection near $C_{\rm eff}\simeq1$ in shells with coupled heat and composition fluxes at the inner boundary. Whether this regime change is connected to the switch of the critical wave at onset remains open.

The nonlinear result needs more before it can be tested in shells. For the pair of distinct azimuthal modes considered here, the quadratic interaction generates no zonal velocity, so the cubic self-saturation is supplied by the mean scalar profiles (Appendix~\ref{sec:radial-validity}); at larger amplitude the annulus equations admit mean zonal flows as well. In shells, Reynolds and thermal stresses drive zonal flows that can make the onset subcritical \citep{Skene2025}. If one self coefficient becomes negative, the interaction enters the difficult case of the Hopf--Hopf interaction, where a modulated state with periodic intensities becomes possible; deciding it requires fifth-order coefficients \citep[\S~8.6.3, remark~2]{Kuznetsov2023}. Testing the coexistence criterion in shells therefore requires spherical eigenfunctions, their scalar feedback and their zonal flow. The factorisation \eqref{eq:factorG}--\eqref{eq:detW} provides the benchmark for such a test.

\section{Conclusions}\label{sec:conclusions}
In the rapidly rotating annulus, when two buoyancies with different diffusivities drive the fluid, the neutral curves have an envelope that, with the diffusivity ratio as a parameter, is a wave front, and the onset threshold is its exposed part. Its swallowtail organises the passage from the straight codensity diagonal to a kinked threshold, at which a broad, fast thermal wave and a narrow, slow compositional wave become neutral together. The swallowtail locus is a unique, increasing algebraic branch $\tau_{\rm sw}(\Pran)$, along which numerical continuation traces the transition from smooth to kinked onset; a kink requires a Lewis number above a critical value between $5.26$ and $6.16$. Hidden arcs of the same front mark where unstable wave-number bands are born and merge, and the compositional end of the threshold shows that heat can stabilise the compositional wave through the phase of its response.

Every neutral root is simple, so these singularities come from the selection of the wave number, not from coalescing eigenvalues. The onset is the zero level of $\max_l\sigma(l;R,S,\tau)$ \citep{Bryzgalova1977}: singularities of matrix families at fixed wave number \citep{Arnold1971} and singularities of the envelope over the wave number \citep{Arnold1990} are distinct sources, and the kink belongs to the second.

The same geometry decides the nonlinear interaction. The coupling of the two waves through the mean scalar profiles factorises into the normals of the two neutral arcs and the responses of the two scalars, and the resulting margin is explicitly positive at every corner of the leading front: wherever the threshold kinks, the mean profiles favour coexistence. The margin closes linearly in $\tau_{\rm sw}-\tau$ at fixed $\Pran$, quadratically in the kink angle. The complete cubic coefficients preserve coexistence at a reference corner and predict a narrow bistable interval near the merger, and simulations confirm the coexisting waves and their invasion thresholds. In shells, zonal flows add a feedback that is absent from the quadratic response of the pair considered here, and spherical eigenfunctions will decide whether the wedge survives; the annulus factorisation is the benchmark for that test.

Our adaptation of Newton's anagram asks for the envelope of a family of curves, and conversely. Here the envelope of the neutral curves gives the threshold, and, conversely, its geometry at the kink decides, through the mean scalar profiles, which waves survive.

\begin{acknowledgments}
This work was supported by the British Academy (Pump Priming Collaboration between UK and EU Partners 2026, award PPHE26\textbackslash 100263). The author acknowledges support of the Institut Henri Poincar\'e (UAR 839 CNRS-Sorbonne Universit\'e), and LabEx CARMIN (ANR-10-LABX-59-01). The author used ChatGPT (OpenAI, GPT-6.1 Sol) and Claude (Anthropic, Claude Opus 5.5) to assist with drafting and editing the text; their use in the computations is described in Appendix~\ref{sec:numerics}. The author takes full responsibility for the content. The author reports no conflict of interest.
\end{acknowledgments}

\section*{Data availability}
The code and data that reproduce the figures, tables and numerical results reported here are available from the author upon request. The spherical comparison uses points read from Fig.~1(b) of \citet{Mather2021}.

\appendix

\begin{table}[!htbp]
\caption{Symbols introduced in the derivations and where they are defined; each is defined again where it first enters.}\label{tab:notation}
\begin{ruledtabular}
\begin{tabular}{lp{.80\textwidth}}
Symbols & Meaning and location\\
\hline
$Z_j=X_j^2$, $z=Z/\tau$ & Squared dimensionless frequency parameter and its rescaling in the cusp quintic (Appendix~\ref{sec:sw-locus}).\\
$J_{\rm unf}$, $B_\tau$ & Growth-unfolding Jacobian and the minor for fixed-$\tau$ sections (Appendix~\ref{sec:leading-surface}).\\
$P_5$, $F_5$ & Cusp quintic in $z$ and its form in $Z$ (Appendices~\ref{sec:sw-locus}, \ref{sec:margin-coefficient}).\\
$\rho_j$, $\gamma_j$ & Weighted growth-sensitivity and scalar-response ratios (Sec.~\ref{sec:criterion}).\\
$u_j$, $v_j$ & Weighted sensitivities $R\partial_R\mu_j$, $S\partial_S\mu_j$, used without division (Sec.~\ref{sec:criterion}).\\
$\mathcal M^{(0)}$, $C_{\rm m}^{\rm lead}$ & Mean-profile coexistence margin and its leading closing coefficient (Appendix~\ref{sec:margin-coefficient}).\\
$C_{\rm m}(\Pran,\beta)$ & Closing coefficient at finite rotation (Sec.~\ref{sec:finite}, Appendix~\ref{sec:margin-coefficient}).\\
$h_Z$ & Half the separation of the squared frequency parameters, $(Z_T-Z_C)/2$ (Appendix~\ref{sec:margin-coefficient}).\\
$\mathscr N$, $A_N$ & Exact neutral polynomial and its control-derivative matrix (Appendices~\ref{sec:sw-locus}, \ref{sec:margin-coefficient}).\\
$\gamma_{\rm gap}$ & Decay gap of noncritical admissible modes (Appendix~\ref{sec:radial-validity}).\\
\end{tabular}
\end{ruledtabular}
\end{table}

\section{Conventions and notation}\label{sec:conventions}
We use the annulus equations and boundary conditions of \citet[\S2]{Simitev2011}, who follows \citet{Busse2002}, with the stream-function and Jacobian conventions used here. In Simitev's notation $P=\Pran$, $L=1/\tau$, $R_t=R$, $R_s=\tau S$ and $\eta^*=\beta$, where $g=\Omega^2s_0$ is the centrifugal acceleration at the mean radius, $R=g\alpha_T\Delta T\,d^3/(\nu\kappa_T)$ and $S=g\alpha_C\Delta C\,d^3/(\nu\kappa_C)$ are positive for destabilising contrasts $\Delta T$ and $\Delta C$ across the gap. With Simitev's Jacobian
\begin{equation*}
J_\psi=(\partial_y\psi)\partial_x-(\partial_x\psi)\partial_y,
\end{equation*}
the substitution $x=x_{\rm S}+\tfrac12$ and
\begin{equation*}
(\psi,\Theta,\Xi)=-(\psi,\Theta,\Gamma)_{\rm S}
\end{equation*}
maps equations (3) of \citet{Simitev2011} onto equation~\eqref{eq:nl}.

Both scalar deviations are measured in units of $\Pran$ times their imposed contrasts, and all rates use the viscous time $d^2/\nu$. The dispersion cubic is equation~\eqref{eq:cubic}, which is $\tau$ times the cubic in equation (C12) of \citet{Monville2019}, so the two dispersion relations have the same roots.

Modes are written as $\sin(n\pi x)e^{\lambda t+\ii ly}$ with $H=(n\pi)^2+l^2$. For $l>0$, a neutral wave $\lambda=\ii\omega$ with $\omega<0$ drifts progradely with phase speed $-\omega/l$. The complex amplitude $A_j$ is the positive Fourier coefficient of the stream function, so a single real Fourier mode has peak amplitude $2|A_j|$.
Table~\ref{tab:notation} lists the symbols introduced in the derivations below.

\section{Exact neutral family and selection of onset}\label{app:neutral}

\subsection{Exact neutral family and its quartic}\label{sec:quartic}
The scalar amplitudes are eliminated in the rational function $\mathcal F$ of equation~\eqref{eq:cubic}. Its eliminated factors vanish only on the negative real axis, so the cubic and rational formulations are equivalent for neutral and growing modes. With $\lambda=\ii HX/\Pran$ and
\begin{equation}
 \mathcal A(X)=\frac{R}{1+X^2}+\frac{\tau^2S}{\tau^2+X^2},\qquad
 \mathcal B(X)=\frac{R}{1+X^2}+\frac{\tau S}{\tau^2+X^2},
 \label{eq:AB-definition}
\end{equation}
the real and imaginary parts of $\mathcal F=0$ read
\begin{equation}
 \mathcal A=\frac{H^3}{l^2},\qquad
 \mathcal B=-\frac{\mathcal A}{\Pran}-\frac{\beta H}{lX}.
 \label{eq:neutral-AB}
\end{equation}
Being linear in $(R,S)$, they give equation~\eqref{eq:family}. At fixed $l$, varying $X$ traces a neutral curve. In the quadrant $R,S>0$ the allowed interval is $-\beta l/(aH^2)<X<-\tau\beta l/(bH^2)$, with $a=1+1/\Pran$ and $b=1+\tau/\Pran$.

The exact envelope $\partial(R,S)/\partial(X,l)=0$ is the quartic
\begin{equation}
 \mathcal E_4(X,l)=l^2X^4+q_3X^3+q_2X^2+q_1X+q_0=0,
 \label{eq:quartic}
\end{equation}
where
\begin{align}
 q_3&=\frac{2(1+\tau)abH^2(3l^2-H)}{\beta l},\qquad
 q_2=d_8l^2-d_3H,\qquad
 d_8=3\tau^2+8\tau+3+\frac{8\tau(1+\tau)}{\Pran},\nonumber\\
 d_3&=\tau^2+3\tau+1+\frac{3\tau(1+\tau)}{\Pran},\qquad
 q_1=\frac{\beta\tau(1+\tau)l(2l^2-H)}{H^2},\qquad
 q_0=\tau^2(H-3l^2).
 \label{eq:quartic-coefficients}
\end{align}
Direct differentiation gives
\begin{equation}
 \frac{\partial(R,S)}{\partial(X,l)}
 =-\frac{2\beta H^3}{\tau(\tau-1)l^4X^2}\,\mathcal E_4.
 \label{eq:quartic-identity}
\end{equation}
Solving the quartic for real $X$ at successive $l$ therefore traces the exact envelope without minimising a neutral threshold. At a self-intersection both pairs $(X,l)$ are kept: they describe different neutral waves at the same $(R,S)$.

\subsection{Neutral roots and the resultant}\label{app:resultant}
The resultant of the real and imaginary parts of $f(\mathrm{i}\omega)$ vanishes whenever the two polynomials share a root, which could a priori be non-real. In the quadrant $R,S>0$ this does not happen. With $p=\Pran$, $a_0=p+1+\tau$, $\varkappa=\beta l/H^2$, $r_*=Rl^2/H^3$ and $s_*=Sl^2/H^3$, division of $f(\mathrm{i}HX/p)$ by $H^4$ gives the real and imaginary parts
\begin{align*}
 P_{\rm r}(X)&=-\frac{a_0}{p}X^2-\varkappa(1+\tau)X+\tau(1-r_*-s_*),\\
 P_{\rm i}(X)&=-\frac{X^3}{p}-\varkappa X^2+\Bigl(1+\tau+\frac{\tau}{p}-r_*-\tau s_*\Bigr)X+\varkappa\tau .
\end{align*}
If $P_{\rm r}$ and $P_{\rm i}$ had a common non-real root, its complex conjugate would also be common, and the quadratic $P_{\rm r}$ would divide the cubic $P_{\rm i}$. Comparing the leading coefficients gives the quotient $X/a_0+\varkappa p^2/a_0^2$, and comparing the constant terms then requires $1-r_*-s_*=a_0^2/p^2$. Since $a_0>p$, the right side exceeds one, whereas $r_*,s_*>0$; this is impossible. Hence, for $R,S>0$, $D_3=0$ holds exactly when $f$ has a neutral root $\mathrm{i}\omega$ with real $\omega$. The same argument excludes two distinct neutral frequencies at one $l>0$, since two distinct real common roots would again make $P_{\rm r}$ divide $P_{\rm i}$.

\subsection{Selecting the physical boundary}\label{sec:selection-boundary}
For fixed $l$, all three Bilharz minors must be positive for strict stability \citep{Bilharz1944,Kirillov2021}, and the stable set of the full problem is their intersection over admissible $l$. Lower-minor tests at an envelope's own wave number therefore cannot discard hidden arcs. Each real root of the quartic is substituted into equation~\eqref{eq:family} with its frequency and wave number retained, and the remaining admissible modes at those controls select the exposed boundary. A later neutral maximum creates another unstable band, and a neutral minimum merges two bands. In a periodic cell, $l=m\Delta l$ replaces the continuous interval.

The potentially unstable wave numbers lie in a bounded set. Let $\mathcal R_+=\max(R,0)+\max(S,0)$. For any root with $\sigma=\operatorname{Re}\lambda\ge0$, the real part of $\mathcal F=0$ gives
\begin{equation}
 H(\sigma+H)\le l^2\left[\frac{\max(R,0)}{\Pran\sigma+H}
 +\frac{\tau\max(S,0)}{\Pran\sigma+\tau H}\right]
 \le\frac{l^2\mathcal R_+}{H},
 \label{eq:real-part-bound}
\end{equation}
because the response denominators have positive real parts and the rotational term is imaginary. Hence $H^3\le l^2\mathcal R_+$, and minimising $H^3/l^2$ at $l^2=(n\pi)^2/2$ gives
\begin{equation}
 0<l\le\mathcal R_+^{1/4},\qquad
 1\le n\le\frac1\pi\left(\frac{4\mathcal R_+}{27}\right)^{1/4}.
 \label{eq:finite-mode-bound}
\end{equation}
The $l=0$ modes are diffusive and stable. These bounds hold at finite rotation and for either sign of the two Rayleigh numbers, so only bounded intervals and finitely many radial indices need testing.

\section{Swallowtail certificate and algebraic locus}\label{app:swallowtail}

\subsection{Leading surface, corners and the swallowtail certificate}\label{sec:leading-surface}
For the leading rapid-rotation surface put $\epsilon=k^2\beta^{-2/3}$, $l=\beta^{1/3}\ell$ and $Y=-X$. Before setting $\epsilon$ to zero, the exact scaled neutral family is
\begin{align}
 \hat R_\epsilon&=
 \frac{(\ell^2+\epsilon)(Y^2+1)
           [b(\ell^2+\epsilon)^2Y-\tau\ell]}
      {(1-\tau)\ell^2Y},\qquad
 \hat S_\epsilon=
 \frac{(\ell^2+\epsilon)(Y^2+\tau^2)
           [\ell-a(\ell^2+\epsilon)^2Y]}
      {(1-\tau)\tau\ell^2Y},
 \label{eq:finite-scaled-family}
\end{align}
and $\epsilon=0$ gives equation~\eqref{eq:lead}. With $w=\ell^3Y$ and $Z=Y^2$, the Jacobian identity \eqref{eq:radical-certificate} of Proposition~\ref{prop:radical} gives the complete leading envelope. A cusp is a zero of both $\rd\hat R/\rd w$ and $\rd\hat S/\rd w$. A corner has two preimages $w_T\ne w_C$ with identical controls, and it is exposed when all remaining admissible modes are stable there.

At finite rotation the reference corner follows from the exact family. For $\Pran=0.3$, $\tau=0.1$ and $\beta=3.786\times10^6$,
\begin{align}
 (R,S)&=(1.126123\times10^{8},\ 7.110042\times10^{8}),\nonumber\\
 (l_T,l_C)&=(88.7884,\ 131.9327),\qquad
 (\omega_T,\omega_C)=(-26675.5,\ -9483.58),\nonumber\\
 (R/R_0,S/S_0)&=(0.713411,\ 0.935654),
 \label{eq:reference-corner}
\end{align}
where the last line uses the leading pure thresholds~\eqref{eq:pure}. The growth rate maximised over the three roots has two distinct, simple neutral maxima at these wave numbers. Their scaled values $(\ell,X)=(0.5697,-1.0139)$ and $(0.8465,-0.1634)$ differ from the leading preimages by $0.07\%$ and $0.03\%$ in $\ell$, and the corner differs from the leading corner $(0.713021,0.935509)$ by $0.055\%$ and $0.016\%$.

At $\Pran=0.3$ the cusp edges and the corner locus meet at the leading swallowtail point
\begin{equation*}
(R/R_0,S/S_0,\tau)=(0.573760,0.832261,0.170059).
\end{equation*}
At $\beta=3.786\times10^6$ the finite-rotation merger occurs at $\tau_c=0.1700462$ (Appendix~\ref{sec:sw-locus}).

At the algebraic point of Appendix~\ref{sec:sw-locus}, the scaled growth $F=\hat\sigma(\ell;\hat R,\hat S,\tau)$ has a quartic maximum at $\ell=0.6731666762$:
\begin{equation*}
F=F_\ell=F_{\ell\ell}=F_{\ell\ell\ell}=0,
\end{equation*}
while $F_{\ell\ell\ell\ell}=-3890.2823$. Implicit differentiation of the cubic gives
\begin{equation}
 J_{\rm unf}=\frac{\partial(F,F_\ell,F_{\ell\ell})}
 {\partial(\log\hat R,\log\hat S,\log\tau)},\qquad
 \det J_{\rm unf}=-61.6393,\qquad
 B_\tau=\det\frac{\partial(F,F_\ell)}{\partial(\log\hat R,\log\hat S)}=2.27083.
 \label{eq:sw-time-minor}
\end{equation}
The full determinant verifies the versal unfolding. The additional minor $B_\tau\ne0$ verifies that the physical slices $\tau=\mathrm{const}$ are generic: the derivative of $\log\tau$ with respect to the quadratic unfolding coefficient is proportional to $B_\tau/\det J_{\rm unf}$. This is the time-function condition in \citet[\S~3.3]{Arnold1990}; here diffusivity ratio plays the role of time in a family of fronts.

The same simple-root derivative calculation has been repeated at all $1199$ finite-$\Pran$ points of the saved swallowtail locus, refining each point with $55$-digit arithmetic. Throughout the sampled range $0.001309\le\Pran\le763.94$, $F_{\ell\ell\ell\ell}<0$, $\det J_{\rm unf}<0$ and $B_\tau>0$. After each row is divided by its Euclidean norm, the minimum absolute determinant is $0.141566$ for $J_{\rm unf}$ and $0.545148$ for the two-row, two-column matrix defining $B_\tau$. The pointwise derivatives and these normalised determinants are available from the author on request. These are numerical checks at the saved points, not a proof excluding an $A_4$ point or a degenerate section between samples or at either Prandtl-number endpoint.

\subsection{Algebraic swallowtail locus and its Prandtl-number limits}\label{sec:sw-locus}
At the swallowtail two cusps of the leading front merge. The cusp condition and its merger are polynomial equations in the front variables $(w,Z)$, \eqref{eq:sw-elimination}--\eqref{eq:sw-algebraic-system}. Eliminating $w$ and $Z$ leaves one equation $\Phi=0$ for $\tau_{\rm sw}(\Pran)$, and the leading coefficients of $\Phi$ give its Prandtl-number limits. These limits are taken after the rapid-rotation limit at fixed positive $\Pran$ and $\tau$. At finite $\beta$ their accuracy need not be uniform as either material parameter approaches an endpoint.

Write $p=\Pran$ temporarily, retain $a=1+1/p$ and $b=1+\tau/p$, and define
\begin{equation}
 A_e=4(1+\tau)ab,\qquad
 B_e=(2+\tau)(1+2\tau)+\frac{5\tau(1+\tau)}p,\qquad
 C_e=\tau(1+\tau).
 \label{eq:sw-elimination-coefficients}
\end{equation}
The radical equation and the cusp condition are the two polynomial equations
\begin{align}
 E(w,Z)&=wZ^2-(A_ew^2-B_ew+C_e)Z-2\tau^2w=0,\nonumber\\
 K(w,Z)&=w(4bw-\tau)(Z+1)(Z^2+2\tau^2)
 +(bw-\tau)(Z-2)Z(A_ew^2-C_e)=0.
 \label{eq:sw-elimination}
\end{align}
On the positive radical sheet $E_Z>0$. Along $E=0$ the cusp condition is $K=0$, and two cusps merge when its derivative along that sheet also vanishes:
\begin{equation}
 E=K=E_ZK_w-E_wK_Z=0,\qquad
 Z>0,\qquad \frac{\tau}{b}<w<\frac1a,\qquad 0<\tau<1.
 \label{eq:sw-algebraic-system}
\end{equation}
Equation~\eqref{eq:radical} then recovers both Rayleigh numbers, the wave number and the frequency, and the certificate of Appendix~\ref{sec:leading-surface} distinguishes a swallowtail from a higher degeneration.

Eliminating $Z$ from $E=K=0$ leaves a quintic in $w$, after removing endpoint and denominator factors. Its discriminant has a factor $\Pi_{14}(\tau;p)$ of degree fourteen in $\tau$ and six in $p$. In the variables
\begin{equation}
 Q=\frac{(1+\tau)^2}{\tau},\qquad
 V=\frac{1+\tau}{p},\qquad
 \Pi_{14}(\tau;p)=\frac{p^6\tau^8}{(1+\tau)^2}\Phi(Q,V),
 \label{eq:sw-invariant-polynomial}
\end{equation}
both unchanged by the scalar exchange $(p,\tau)\mapsto(p/\tau,1/\tau)$,
\begin{align}
 \Phi(Q,V)={}&Q(Q+2)\varphi_0(Q)+2QV\varphi_1(Q)+3V^2\varphi_2(Q)
             +2V^3\varphi_3(Q)+V^4\varphi_4(Q)\nonumber\\
 &+1125V^5[V+2(Q+1)](Q^2-4Q-36),
 \label{eq:sw-Phi}\\
 \varphi_0(Q)={}&64Q^6-704Q^5+1744Q^4+2168Q^3
          -14506Q^2-31066Q-21609,\nonumber\\
 \varphi_1(Q)={}&224Q^6-2032Q^5+2756Q^4+8236Q^3
          -73883Q^2-126119Q-33957,\nonumber\\
 \varphi_2(Q)={}&368Q^6-2216Q^5-4140Q^4+3100Q^3
          -116815Q^2-131586Q-18522,\nonumber\\
 \varphi_3(Q)={}&1100Q^5-5628Q^4-21366Q^3-5822Q^2
          -328599Q-7938,\nonumber\\
 \varphi_4(Q)={}&2686Q^4-10238Q^3-82331Q^2-83772Q-371304.
 \label{eq:sw-coefficient-polynomials}
\end{align}
The prefactor in \eqref{eq:sw-invariant-polynomial} is positive for $p,\tau>0$, so $\Phi$ and $\Pi_{14}$ give the same locus. A real root must still satisfy \eqref{eq:sw-algebraic-system} on the positive sheet. For $p=3/10$, selection gives $\tau_{\rm sw}=0.1700590$ and $(R/R_0,S/S_0)=(0.573760,0.832261)$, independently of the derivative certificate.

\paragraph{A cusp quintic in frequency.}
For the nonlinear margin below it is useful to eliminate $w$ instead of $Z$ from \eqref{eq:sw-elimination}. Removing endpoint and denominator factors leaves a quintic in $z=Z/\tau$:
\begin{align}
 P_5(z;Q,V)&=\sum_{j=0}^5 d_j(Q,V)z^j,\label{eq:cusp-frequency-quintic}\\
 d_5&=5(Q+2+4V),\qquad d_0=12,\nonumber\\
 d_4&=4Q^2-18Q-7-2V(Q-20)+18V^2,\nonumber\\
 d_3&=-[4Q^3-20Q^2+25Q-12+2V(2Q^2+Q-15)+9V^2(Q-2)],\nonumber\\
 d_2&=28Q^2-72Q-31+10V(4Q+1)+45V^2,\nonumber\\
 d_1&=-4(13Q-1+25V).\nonumber
\end{align}
Its discriminant contains the same factor $\Phi(Q,V)$ as \eqref{eq:sw-Phi}. The physical swallowtail is a positive double root $P_5=P_{5,z}=0$, $P_{5,zz}\ne0$, selected by \eqref{eq:sw-algebraic-system}. The other factors of the discriminant do not replace this sheet selection. At an ordinary double root, the linear subresultant of $P_5$ and $P_{5,z}$ gives $z$ as a rational function of $Q,V$ on the discriminant; keeping the quintic is a shorter representation.

\paragraph{Small Prandtl number.} As $p\to0$ with $\tau>0$, $V\to\infty$, and the coefficient of $V^6$ in $\Phi$ is $1125(Q^2-4Q-36)$. Equivalently,
\begin{equation}
 \Pi_{14}(\tau;0)
 =1125\tau^6(1+\tau)^4
 (\tau^2+6\tau-1)(\tau^2-6\tau-1),
 \label{eq:sw-small-Pr-factor}
\end{equation}
whose only root in $0<\tau<1$ gives
\begin{equation}
 \tau_0=\sqrt{10}-3,\qquad
 \tau_{\rm sw}(p)=\tau_0+\frac{9(10-\sqrt{10})}{200}p^2+O(p^3);
 \label{eq:sw-small-Pr}
\end{equation}
the linear term vanishes. In terms of the Lewis number $\mathit{Le}=1/\tau$,
\begin{equation*}
\mathit{Le}_{\rm sw}\to\sqrt{10}+3=6.1623.
\end{equation*}
\paragraph{Large Prandtl number.} As $p\to\infty$, $V\to0$ and $\Phi(Q,0)=Q(Q+2)\varphi_0(Q)$. Positive $\tau$ excludes the first two factors, so the limit is the sextic $\varphi_0(Q_\infty)=0$ with $Q_\infty>4$. A Sturm count gives exactly one root in that interval:
\begin{equation}
 Q_\infty=7.453608,\qquad
 \tau_\infty=
 \frac{2}{Q_\infty-2+\sqrt{Q_\infty(Q_\infty-4)}}
 =0.1899831.
 \label{eq:sw-large-Pr}
\end{equation}
Written in $\tau$, the same condition is the reciprocal degree-twelve equation $\tau^6\varphi_0((1+\tau)^2/\tau)=0$. Implicit differentiation gives the approach to the plateau,
\begin{equation}
 \tau_{\rm sw}(p)=\tau_\infty-\frac{C_\infty}{p}+O(p^{-2}),
 \qquad
 C_\infty=
 \frac{(1+\tau_\infty)\Phi_V(Q_\infty,0)}
 {(1-\tau_\infty^{-2})\Phi_Q(Q_\infty,0)}
 =0.0145300>0.
 \label{eq:sw-large-Pr-correction}
\end{equation}
Sturm isolation gives $7.4536<Q_\infty<7.4537$ with $\varphi_0'(Q_\infty)>0$ and $\varphi_1(Q_\infty)<0$. Since $\tau_\infty<1$, these signs prove $C_\infty>0$, so the branch approaches its large-$p$ endpoint from below. In terms of the Lewis number,
\begin{equation*}
\mathit{Le}_{\rm sw}\to1/\tau_\infty=5.2636.
\end{equation*}
\paragraph{Finite rotation.} The finite-rotation singularity is also algebraic. With $L=l^2$ and $H=k^2+L$, the determinant $D_3$ of Sec.~\ref{sec:envelope} makes $\mathscr N=D_3/(p^4\tau H^5)$ a degree-ten polynomial in $L$. At a regular real-frequency neutral sheet with a simple temporal eigenvalue, the swallowtail satisfies
\begin{equation}
 \mathscr N=\partial_L\mathscr N
 =\partial_L^2\mathscr N=\partial_L^3\mathscr N=0
 \label{eq:sw-finite-system}
\end{equation}
for the unknowns $(R,S,\tau,L)$ at prescribed $p$, $\beta$ and $k$. The quadruple root is a root in the wave number, while the temporal eigenvalue stays simple. For $p=0.3$ and $\epsilon=k^2\beta^{-2/3}$,
\begin{equation*}
\tau_c=0.1700590-0.03132\,\epsilon+O(\epsilon^2).
\end{equation*}
At $k=\pi$, $\beta=3.786\times10^6$ the exact finite-rotation merger is $\tau_c=0.1700462$, $(R/R_0,S/S_0)=(0.574123,0.832401)$ and $l_{\rm sw}=104.863$, with $S_0$ evaluated at that $\tau$ by equation~\eqref{eq:pure}.

\paragraph{The complete locus.} Figure~\ref{fig:sw-locus} follows the branch over $0\le\Pran\le\infty$ in the compactified coordinate $\Pran=\tan\theta$. Its endpoints are \eqref{eq:sw-small-Pr} and \eqref{eq:sw-large-Pr}, and in $\theta$ the right endpoint has finite slope $C_\infty$. The computed continuation has $1199$ finite points and increases throughout, as the following exact result requires.

\paragraph{Monotonicity of the algebraic branch.} Write the polynomial as a series in $V$,
\begin{equation}
 \Phi=\sum_{j=0}^{6}\Phi_j(Q)\,V^j,
 \label{eq:phi-series}
\end{equation}
with the coefficients
\begin{gather*}
 \Phi_0=Q(Q+2)\varphi_0,\quad \Phi_1=2Q\varphi_1,\quad \Phi_2=3\varphi_2,\quad \Phi_3=2\varphi_3,\quad \Phi_4=\varphi_4,\\
 \Phi_5=2250(Q+1)(Q^2-4Q-36),\qquad \Phi_6=1125(Q^2-4Q-36).
\end{gather*}
Exact rational arithmetic establishes three facts. First, the coefficients of $\Phi_j(4+x)$ change sign exactly once, so each $\Phi_j$ has exactly one root $Q_j>4$, where it changes from negative to positive. Second, all coefficients of $\Phi_j'(7+x)$ are positive. Third, the roots are ordered,
\begin{equation}
 7<Q_0<Q_1<\dots<Q_5=Q_6<8.33,\qquad Q_0=7.45361,\qquad Q_6=2+2\sqrt{10}=8.32456.
 \label{eq:phi-roots}
\end{equation}
For $V>0$ all $\Phi_j$ have one sign outside $(Q_0,Q_6)$, while inside this interval $\Phi_Q>0$, so $\Phi=0$ has exactly one root $Q(V)>4$. At that root the non-zero $\Phi_j(Q)$ change from positive to negative as $j$ increases. With $k$ separating the two groups, and because $\Phi=0$,
\begin{equation}
 V\Phi_V=\sum_{j}(j-k)\,\Phi_jV^j<0,\qquad\text{hence}\qquad \frac{\rd Q}{\rd V}=-\frac{\Phi_V}{\Phi_Q}>0.
 \label{eq:phi-monotone}
\end{equation}
Since $Q=(1+\tau)^2/\tau$ decreases with $\tau$ on $(0,1)$ and $\Pran=(1+\tau)/V$, the algebraic branch $\tau_{\rm sw}(\Pran)$ increases strictly from $\sqrt{10}-3$ to $\tau_\infty$. This result concerns the algebraic branch; that its points are swallowtails of the exposed front is established by the certificate at $\Pran=0.3$ and sampled along the branch in Appendix~\ref{sec:leading-surface}.

\section{Details of the cubic interaction}\label{app:cubic}
The mean profiles of equation~\eqref{eq:means} follow from the quadratic scalar equations. For the positive Fourier components, $\hat\Theta_k=-\ii l_kT_{T,k}A_k$ and $\hat\Xi_k=-\ii l_kT_{C,k}A_k$, and averaging a wave and its conjugate over their phase gives
\begin{equation}
 \langle J(\psi_k,q_k)\rangle
 =2\pi l_k^2\operatorname{Re}(T_{q,k})|A_k|^2\sin(2\pi x),
 \qquad (q,T_{q,k})=(\Theta,T_{T,k})\ \text{or}\ (\Xi,T_{C,k}).
 \label{eq:nf-mean-forcing}
\end{equation}
The zero-frequency balances
\begin{equation*}
\bar\Theta_k''=\Pran\,\langle J(\psi_k,\Theta_k)\rangle
\end{equation*}
and
\begin{equation*}
\tau\bar\Xi_k''=\Pran\,\langle J(\psi_k,\Xi_k)\rangle,
\end{equation*}
with $\operatorname{Re}T_{T,k}=H_k|T_{T,k}|^2$ and
\begin{equation*}
\operatorname{Re}T_{C,k}=\tau H_k|T_{C,k}|^2,
\end{equation*}
then give equation~\eqref{eq:means}, and the factor $\tau$ cancels in the compositional balance.

\subsection{Projection of the cubic coefficients}\label{app:projection}
The cubic interaction combines feedback through the mean scalar profiles with the sum and difference harmonics. We compute these contributions separately, so the geometric factorisation can be compared with the complete interaction. Throughout, $H_j=\pi^2+l_j^2$, $T_{T,j}=(\ii\Pran\,\omega_j+H_j)^{-1}$,
\begin{equation*}
T_{C,j}=(\ii\Pran\,\omega_j+\tau H_j)^{-1}
\end{equation*}
and
\begin{equation*}
\mathcal F_{\lambda,j}=H_j+\Pran\,l_j^2(RT_{T,j}^2+\tau ST_{C,j}^2).
\end{equation*}

Write
\begin{equation*}
\boldsymbol\phi_j=\boldsymbol e_j\sin(\pi x)\exp[\ii(l_jy+\omega_jt)]
\end{equation*}
with unit stream-function coefficient. The bracket $[\,\cdot\,]_n$ denotes the coefficient of $\sin(n\pi x)$ at the indicated wave number and frequency. For radial index $n$, azimuthal wave number $q$ and frequency $\omega'$, put $H_{nq}=(n\pi)^2+q^2$. The modal pencil $\mathsf L(q,\omega',n)$ of the unmultiplied vorticity equation is the main matrix \eqref{eq:M} at $\lambda=\ii\omega'$, wave number $q$ and radial index $n$, with the sign of its first row reversed. The right vector and a left nullvector are
\begin{equation}
 \boldsymbol e_j=(1,-\ii l_jT_{T,j},-\ii l_jT_{C,j})^{\mathsf T},\qquad
 (\boldsymbol e^{\rm L}_j)^{\mathsf T}=(1,-\ii l_jRT_{T,j},-\ii l_j\tau ST_{C,j}).
 \label{eq:nf-vectors}
\end{equation}
The time-derivative matrix is
\begin{equation*}
\mathsf E_j=\operatorname{diag}(-H_j,\Pran,\Pran),
\end{equation*}
and
\begin{equation}
 (\boldsymbol e^{\rm L}_j)^{\mathsf T}\mathsf E_j\boldsymbol e_j=-\mathcal F_{\lambda,j}
 \label{eq:nf-projection-denominator}
\end{equation}
fixes the amplitude normalisation and the sign of the projection.

For vector fields $\boldsymbol a=(\psi_a,\Theta_a,\Xi_a)$ and $\boldsymbol b$, write the quadratic advection as
\begin{equation}
 \mathcal N(\boldsymbol a,\boldsymbol b)=
 \begin{pmatrix}
 J(\psi_a,\nabla^2\psi_b)\\
 \Pran\,J(\psi_a,\Theta_b)\\
 \Pran\,J(\psi_a,\Xi_b)
 \end{pmatrix},\qquad
 \mathcal B(\boldsymbol a,\boldsymbol b)
 =\mathcal N(\boldsymbol a,\boldsymbol b)+\mathcal N(\boldsymbol b,\boldsymbol a).
 \label{eq:nf-bilinear}
\end{equation}
Let $\boldsymbol h^0_k$ be the response to $|A_k|^2$, and $\boldsymbol h^\mp_{jk}$ the responses to $A_j\overline A_k$ and $A_jA_k$. Their radial-index-two coefficients, obtained with the modal matrix $\mathsf L$ defined above, are
\begin{align}
 \boldsymbol h^0_k&=-\mathsf L(0,0,2)^{-1}
                 [\mathcal B(\boldsymbol\phi_k,\overline{\boldsymbol\phi}_k)]_2,
                 \nonumber\\
 \boldsymbol h^\mp_{jk}&=-\mathsf L(l_j\mp l_k,\omega_j\mp\omega_k,2)^{-1}
                 [\mathcal B(\boldsymbol\phi_j,\overline{\boldsymbol\phi}_k)\ \text{or}\ \mathcal B(\boldsymbol\phi_j,\boldsymbol\phi_k)]_2,
                 \qquad j\ne k.
 \label{eq:nf-harmonic-resolvents}
\end{align}
With $\mathcal P_j$ selecting the radial-index-one component at $(l_j,\omega_j)$, the resonant cubic forcing for distinct waves and the complete coefficient are
\begin{equation}
 \boldsymbol r_{jk}=\mathcal P_j\big[\mathcal B(\boldsymbol\phi_j,\boldsymbol h^0_k)
       +\mathcal B(\boldsymbol\phi_k,\boldsymbol h^-_{jk})
       +\mathcal B(\overline{\boldsymbol\phi}_k,\boldsymbol h^+_{jk})\big],\qquad
 c_{jk}=-\frac{(\boldsymbol e^{\rm L}_j)^{\mathsf T}\boldsymbol r_{jk}}
                  {(\boldsymbol e^{\rm L}_j)^{\mathsf T}\mathsf E_j\boldsymbol e_j};
 \label{eq:nf-full-projection}
\end{equation}
the self forcing contains only the first term. The three terms give the mean, difference and sum contributions, and the mean term reduces to equation~\eqref{eq:cjk}. Quadratic products of two critical radial sines have radial index two, and the product of an index-two response with a critical mode contains indices one and three only, so the cubic coefficients close without radial truncation. The implementation evaluates these products exactly and checks them against an independent modal product.

The scalar exchange of Sec.~\ref{sec:model} preserves the nonlinear equations under $(\Theta,\Xi)\mapsto(\tau\Xi,\tau\Theta)$ with $\psi$ fixed. With the wave labels exchanged, $\mathcal W'$ is $\tau^2\mathcal W$ with rows and columns swapped and $G'_{\rm full}$ is $G_{\rm full}$ with rows and columns swapped, so $\chi'=\chi$ and the selection transfers to $\tau>1$.

The pole of $\rho(Z)$ at $Z=2\tau^2$ is a coordinate singularity where $\partial_R\mu=0$. It need not be avoided by all corners. At $\Pran=0.3$, $\tau=0.03$, for example, $Z_C=0.00163085<2\tau^2=0.0018$ and $\rho_C=-29.8516$, while $\mathcal M^{(0)}=0.843105>0$. Here heat reduces the growth of the compositional wave. Thus positivity of the margin does not require both waves to have positive thermal sensitivities, and the two differences in the ratio numerator cannot be interpreted independently of its denominator. The homogeneous form retains the normal orientation without that restriction.

The sign is a property of this neutral family and its scalar feedback. It does not follow from a generic crossing alone: the growth branches $\sigma_T=u-(l-l_T)^2$ and $\sigma_C=u+\alpha v+v^2-(l-l_C)^2$ have $\det J_\mu=\alpha$ at $u=v=0$, even when $l_T\ne l_C$. Harmonic feedback also lies outside the identity and produces the narrow full-cubic bistable interval (Sec.~\ref{sec:merger}).

The map in Fig.~\ref{fig:map} evaluates the exact margin \eqref{eq:margin-exact} at the leading corners of a $110\times121$ grid over $10^{-2}\le\Pran\le10^2$, continued numerically in $\tau$ up to the merger. Next to the merger the two preimages nearly coincide, so the $363$ nodes within a relative distance $10^{-3}$ of $\tau_{\rm sw}$ are solved in $65$-digit arithmetic; all $13\,310$ nodes converge. Independent cubic projections check the identity at selected corners.

\subsection{Algebraic closing coefficient}\label{sec:margin-coefficient}
Write $p=\Pran$ in this subsection. The normal form \eqref{eq:sw-section-normal-form} gives a square-root separation of the two waves. Both the neutral-arc normals and the scalar-response columns differ linearly in that separation. Their determinants in equation~\eqref{eq:detG} therefore vanish as $\delta^{1/2}$, with $\delta=\tau_{\rm sw}-\tau$, and the margin is proportional to $\delta$, or to the square of the small kink angle. The proportionality coefficient can be found algebraically at the merger.

The frequency coordinate is regular at merger: at fixed controls, the leading real neutral condition is $\ell^4=\mathcal A(Z)$ with
\begin{equation*}
\mathcal A(Z)=\hat R/(1+Z)+\tau^2\hat S/(\tau^2+Z).
\end{equation*}
Its derivative gives $4\ell^3=\mathcal A'(Z)Z_\ell$, and $\mathcal A'(Z)<0$, so $Z_\ell\ne0$. Let $Z_*=X_*^2$ there, and let $h_Z=(Z_T-Z_C)/2$. Equation~\eqref{eq:margin-exact} gives
\begin{equation}
 \mathcal M^{(0)}\sim
 \frac{3(Z_*^2+2\tau^2)}{(1+Z_*)^2(Z_*+\tau^2)^2}\,h_Z^2.
 \label{eq:margin-local-Z}
\end{equation}
Define $F_5(Z;\tau,p)=\tau^5P_5(Z/\tau;Q,V)$ using \eqref{eq:cusp-frequency-quintic}. At its physical double root, $F_5=F_{5,Z}=0$, the squared half-separation of the two cusp roots is $2\delta F_{5,\tau}/F_{5,ZZ}$. The two corner preimages have three times that value by \eqref{eq:sw-section-normal-form}. Thus
\begin{equation}
 \mathcal M^{(0)}=C_{\rm m}^{\rm lead}\delta+o(\delta),\qquad
 C_{\rm m}^{\rm lead}=
 \left.\frac{18(Z^2+2\tau^2)F_{5,\tau}}
 {(1+Z)^2(Z+\tau^2)^2F_{5,ZZ}}\right|_{\rm sw}.
 \label{eq:C-leading-F5}
\end{equation}
Here $F_{5,\tau}$ is taken at fixed $Z,p$, while $F_{5,ZZ}$ is taken at fixed $\tau,p$. No nearby corners or fitted exponents enter this coefficient. In the invariant variables already used for the swallowtail locus, the explicit form is
\begin{equation}
 C_{\rm m}^{\rm lead}=
 \left.\frac{18(z^2+2)[(\tau^2-1)P_{5,Q}+(\tau^2/p)P_{5,V}]}
 {(1+\tau z)^2(z+\tau)^2P_{5,zz}}\right|_{P_5=P_{5,z}=0}.
 \label{eq:C-leading-invariants}
\end{equation}
The derivatives of $P_5$ hold its other arguments fixed. The coefficient is defined on the physical swallowtail locus, not at independently chosen $p,\tau$. At $p=0.3$, $C_{\rm m}^{\rm lead}=6.98826800991$. Direct corner calculations give $\mathcal M^{(0)}/\delta=6.98732367$, $6.98825856$ and $6.98826792$ at $\delta=10^{-4},10^{-6},10^{-8}$.

The Prandtl-number limits also follow from the quintic. As $p\to0$, $P_5/V^2\to9z^2[2z^2-(Q-2)z+5]$, whose physical double root has $z=\sqrt{10}/2$, $\tau=\sqrt{10}-3$. Along $P_5=0$, $P_{5,V}=O(1)$ while $P_{5,Q}$ and $P_{5,zz}$ are $O(V^2)$, so the $P_{5,V}/p$ term is subleading in the ratio. Substitution gives
\begin{equation}
 \lim_{p\to0}C_{\rm m}^{\rm lead}=\sqrt{10}+\frac{10}{3}
 =6.49561099350\ldots,\qquad
 \lim_{p\to\infty}C_{\rm m}^{\rm lead}=7.22526496039\ldots .
 \label{eq:C-Pr-limits}
\end{equation}
The large-$p$ value is algebraic: set $V=0$ in \eqref{eq:cusp-frequency-quintic} and \eqref{eq:C-leading-invariants}, selecting $z=1.48367546851\ldots$ and the root $\tau_\infty$ of \eqref{eq:sw-large-Pr}. These limits are taken after the rapid-rotation limit.

\paragraph{Finite rotation.}
The same calculation uses the exact neutral polynomial $\mathscr N$ of \eqref{eq:sw-finite-system}, without a thin-column approximation. Write $L=l^2$, retain $H=L+k^2$, and set
\begin{equation}
 A_N=\frac{\partial(\mathscr N,\mathscr N_L,\mathscr N_{LL})}
 {\partial(R,S,\tau)},\qquad
 \boldsymbol v=\frac{L}{H}
 \begin{pmatrix}(1+Z)^{-1}\\(\tau^2+Z)^{-1}\end{pmatrix}.
 \label{eq:C-finite-data}
\end{equation}
At a non-degenerate merger the exact coefficient is
\begin{equation}
 C_{\rm m}(p,\beta)=
 \left.\frac{24RS\det A_N\det(\boldsymbol v,\boldsymbol v_L)}
 {\mathscr N_{LLLL}(R\mathscr N_Rv_1+S\mathscr N_Sv_2)^2}\right|_{\rm sw}.
 \label{eq:C-finite-polynomial}
\end{equation}
All $L$-derivatives hold $R,S,\tau,p,\beta,k$ fixed. Multiplication of $\mathscr N$ by a non-zero smooth factor leaves this expression unchanged: the three rows of its unfolding acquire a triangular transformation with that factor on the diagonal, while the fourth wave-number derivative and the squared denominator acquire the same total power. Locally $\mathscr N$ is such a non-zero multiple of the simple-root growth function, so the formula also follows directly by using $\sigma$ in its place.

For evaluation, a factorised form avoids expansion of the degree-ten polynomial. Define
\begin{align}
 a_0&=p+1+\tau,\quad U=H^3-(R+S)L,\quad
 B_0=[p+\tau(1+p)]H^3-p(R+\tau S)L,\nonumber\\
 V_0&=a_0B_0-p\tau U,\nonumber\\
 W_0&=-p^3\tau U^2+p^2(1+\tau)B_0U
 +p\tau a_0[2a_0-3(1+\tau)]UH^3\nonumber\\
 &\hspace{1em}+(1+\tau)a_0^2B_0H^3-\tau a_0^3H^6,\nonumber\\
 \mathscr N&=HUV_0^2+p\beta^2LW_0+p^4(1+\tau)^2\beta^4L^2H^2.
 \label{eq:neutral-polynomial-factorised}
\end{align}
Combining the real and imaginary neutral cubic equations recovers the frequency parameter as
\begin{equation}
 Z=\frac{p^2\tau^2\beta^2L(a_0^2H^3-p^2U)^2}
 {H^2[a_0HV_0+p^3(1+\tau)\beta^2L]^2}.
 \label{eq:frequency-rational-finite}
\end{equation}
Differentiating this rational expression gives the $Z_L$ needed in $\boldsymbol v_L$. At the stationary neutral root its derivative agrees with that of the temporal branch: the difference of off-neutral extensions is proportional to growth, whose first $L$-derivative vanishes there.

\begin{table}[!htbp]
\centering
\caption{Merger and exact local slope of the mean-profile margin at $\Pran=0.3$, $k=\pi$. The last row is the leading rapid-rotation limit; finite rows use the full neutral polynomial. Nearby-corner solutions independently approach each finite-rotation slope.}\label{tab:C-finite}
\begin{tabular}{ccc}
\toprule
$\beta$ & $\tau_c$ & $C_{\rm m}$\\
\midrule
$10^4$ & 0.169260139430 & 7.052656535\\
$10^5$ & 0.169909847213 & 7.001363983\\
$10^6$ & 0.170027787410 & 6.991053940\\
$3.786\times10^6$ & 0.170046186808 & 6.989412535\\
$\infty$ & 0.170058955446 & 6.988268010\\
\bottomrule
\end{tabular}
\end{table}

The sensitivity ratio also has an exact finite-rotation form. Differentiating the exact neutral family at fixed $l$ gives
\begin{equation*}
\rho=-(\partial_X\log R)/(\partial_X\log S).
\end{equation*}
At a stationary neutral wave the envelope quartic reduces this to
\begin{equation}
 \rho_j=\frac{(a^{\rm rot}_j-Z_j)(Z_j+\tau^2)}
 {(1+Z_j)(Z_j-a^{\rm rot}_j\tau^2)},\qquad
 a^{\rm rot}_j=2-\frac{k^2}{l_j^2}.
 \label{eq:rho-finite}
\end{equation}
Together with equation~\eqref{eq:margin-ratios-main}, this gives the finite-rotation mean-profile margin wherever the ratio coordinates are defined; \eqref{eq:margin-homogeneous} covers their poles. The reduction is checked symbolically using the envelope quartic. The ratio and homogeneous forms also agree with all $85$ independently projected matrices in the stored finite-rotation continuation. These comparisons check the implementation, without extending the leading positivity theorem to all finite rotation rates.

Table~\ref{tab:C-finite} evaluates \eqref{eq:C-finite-polynomial}. For $p=0.3$ and $\epsilon=k^2\beta^{-2/3}$, implicit differentiation of the four merger equations gives
\begin{equation}
 C_{\rm m}(0.3,\beta)=6.98826800991
 +2.81290861319\,\epsilon+O(\epsilon^2).
 \label{eq:C-finite-correction}
\end{equation}
This coefficient concerns the mean-profile margin only. The positive identity and its merger slope do not include the harmonic contribution responsible for the full-cubic bistable interval.

\section{Range of the two-mode reduction}\label{sec:radial-validity}
Radial modes $n\geq2$ lie only $O(\beta^{-2/3})$ above onset \citep{Busse1986}. The $n=2$ thresholds exceed onset by $0.13\%$ (thermal) and $0.06\%$ (compositional) along the ray through the reference corner, control distances rather than error bounds, and the simulations retain $6$--$10$ radial modes (Appendix~\ref{sec:numerics}).

The two-mode reduction requires the remaining modes to decay faster than the two critical amplitudes evolve. The finite-mode bound \eqref{eq:finite-mode-bound} restricts the search for additional instabilities to finitely many radial indices and a bounded interval of wave numbers, where the spectrum is then examined directly. The reduction also needs an isolated critical pair in a prescribed periodic cell and bounded responses at the noncritical combination frequencies.

The radial threshold gaps can be measured without a threshold minimisation. Along the ray $(R,S)=\rho(R_\times,S_\times)$ through the reference corner, solving the exact neutral family with its envelope condition for $(X,l,\rho)$ for each radial index gives
\begin{align}
 \rho_{2,T}-1&=1.2567\times10^{-3},&
 \rho_{2,C}-1&=5.6802\times10^{-4},\nonumber\\
 \rho_{3,T}-1&=3.3699\times10^{-3},&
 \rho_{3,C}-1&=1.5187\times10^{-3}.
 \label{eq:radial-threshold-gaps}
\end{align}
These are relative control distances along this ray and should not be read as nonlinear error bounds. Since the leading thin-column cubic loses its radial dependence, radial isolation cannot be inferred from that approximation alone.

For a discrete isolated pair, define the noncritical decay gap
\begin{equation}
 \gamma_{\rm gap}=-\sup_{\text{noncritical admissible modes}}
           \operatorname{Re}\lambda>0.
 \label{eq:noncritical-gap}
\end{equation}
A sufficient slow-time condition is $\max_j|\mu_j|\ll\gamma_{\rm gap}$, together with bounded noncritical resolvents in the projection. The basic-state scans at the three simulation corners give $\gamma_{\rm gap}=3.29$, and the travelling-wave calculations include all retained radial perturbations. Enlarging the periodic cell admits sidebands arbitrarily close to a critical wave number, which require a modulation or multimode description, so the validation applies to the stated cells.

The self coefficients \eqref{eq:gjj-positive} depend on the saturation mechanism. For a pure $n=1$ wave $J(\psi_j,\nabla^2\psi_j)=0$, and cross products of distinct wave numbers generate no quadratic zonal velocity, so the mean scalar terms give the complete self coefficients at this order of the expansion. In the internally heated spherical cases of \citet{Skene2025}, the zonal flow driven by Reynolds and thermal stresses can make the Hopf bifurcation subcritical, and the coexistence test must then include that contribution.

\section{Numerical method and convergence}\label{sec:numerics}
The spatial calculations test two predictions of the cubic equations: the mixed state and the invasion thresholds. The simulations expand each real field in radial sines and azimuthal Fourier modes,
\begin{equation}
 U(x,y,t)=\sum_{n=1}^{N}\sum_{m=-M}^{M}U_{nm}(t)\sin(n\pi x)e^{\ii m\Delta l y},
 \qquad U_{n,-m}=\overline{U_{nm}},
 \label{eq:galerkin}
\end{equation}
in a periodic cell of length $L_y=2\pi/\Delta l$. In the full annulus $\Delta l=d/s_0$. The amplitude $|\psi_{1m}|$ follows the convention of Appendix~\ref{sec:conventions}. Reference calculations retain $(N,M)=(6,7)$ with $N_x=12$ interior radial and $N_y=24$ azimuthal quadrature points, and refinement uses $(10,10)$ on $(N_x,N_y)=(20,32)$. The conditions $2(N_x+1)>3N$ and $N_y>3M$ prevent quadratic aliasing into retained modes. Each linear modal matrix is exponentiated and advection is treated with fourth-order exponential time differencing \citep{CoxMatthews2002}, with reference step $\Delta t=2\times10^{-4}$ and smaller steps for the rapid-rotation comparisons. Reality, $U_{n,-m}=\overline{U_{nm}}$, is imposed after every step. The tools named in the Acknowledgments, ChatGPT (OpenAI, GPT-6.1 Sol) and Claude (Anthropic, Claude Opus 5.5), assisted in writing the computational and plotting scripts and the symbolic checks of the identities in Appendices~\ref{app:neutral}--\ref{app:cubic}. The author directed this work and verified the derivations and the reported numbers; automated checks reproduce the reported thresholds, coefficients and benchmarks.

Two independent evaluations of the nonlinear term agree: padded quadrature and a sparse convolution give the same advection, which conserves kinetic energy and both scalar variances to roundoff. At the reference mixed state, halving the time step changes amplitudes by $1.94\times10^{-7}$ relatively and refinement changes them by $4.63\times10^{-5}$. At $\beta=10^6$ the corresponding changes are $1.82\times10^{-4}$ and $8.26\times10^{-5}$. These numerical errors are far below the deviations from the cubic predictions examined next.

The thermal, compositional and reference mixed states start from small random perturbations of every retained mode, and the weakly mixed state continues a developed mixed state towards onset. After a perturbation of all modes the solution returns to the mixed state within $5.81\times10^{-5}$. The attracting amplitudes approach the cubic values as the growth rates decrease (Table~\ref{tab:dns}). For invasion, Newton iteration with a phase condition finds each pure travelling wave in its co-moving frame, eigenvalues of the full real Galerkin Jacobian give the invading growth, and continuation locates its zero. These secondary thresholds approach the cubic wedge towards onset, and mixed states at $\beta=10^5$ and $10^6$ stay within $10\%$ of the cubic amplitudes (Table~\ref{tab:validation}). Figure~\ref{fig:fields} shows the developed mixed state.

\end{document}